\documentclass[prd, preprint, tightenlines, a4paper, eqsecnum]{revtex4-1}
\usepackage[latin1]{inputenc}
\pdfoutput=1

\usepackage{amsmath}    % need for subequations
\usepackage{amsfonts}
\usepackage{amssymb}
\usepackage{graphicx}   % need for figures
\usepackage{verbatim}   % useful for program listings
\usepackage{color}      % use if color is used in text
\usepackage{subfigure}  % use for side-by-side figures
\usepackage{hyperref}   % use for hypertext links, including those to external documents and URLs

\begin{document}

%%% AUTHOR

\author{Hugo R. C. Ferreira}
\email{pmxhrf@nottingham.ac.uk}
\affiliation{School of Mathematical Sciences, University of Nottingham, Nottingham NG7 2RD, United Kingdom}

\title{Stability of warped AdS${}_3$ black holes in Topologically Massive Gravity under scalar perturbations}

%%% ABSTRACT

\begin{abstract}
We demonstrate that the warped AdS${}_3$ black hole solutions of Topologically Massive Gravity are classically stable against massive scalar field perturbations by analysing the quasinormal and bound state modes of the scalar field. In particular, it is found that although classical superradiance is present it does not give rise to superradiant instabilities. The stability is shown to persist even when the black hole is enclosed by a stationary mirror with Dirichlet boundary conditions. This is a surprising result in view of the similarity between the causal structure of the warped AdS${}_3$ black hole and the Kerr spacetime in 3+1 dimensions. This work provides the foundations for the study of quantum field theory in this spacetime.
\end{abstract}

%%% KEYWORDS

%\keywords{black hole, stability}

\maketitle

%%%%%%%%%%%%%%%%%%%%%%%%%%%%%%%%%%%%%%%%%%%%%%%%%%%%%%%%%%%%%%%%%%

%%% INTRODUCTION

\section{Introduction}
\label{sec:intro}

In the absence of a quantum theory of gravity there have been many attempts in the last decades to reconcile the physics of gravitational and quantum phenomena. Since such a theory would only manifest itself in extreme situations, such as black holes or the Big Bang, the study of quantum effects in black holes is an important theoretical step in that direction. Indeed, the vast amount of research on quantum effects on black holes, including the famous Hawking effect \cite{Hawking:1974sw}, has been made on the assumption that the spacetime itself remains classical and only matter fields on this background are quantised (see \cite{Birrell1984a,Wald1994b,Parker2009}).

The study of quantum field theory on black hole backgrounds have mostly been restricted to asymptotically flat and asymptotically anti-de Sitter (AdS) spacetimes. This is due to the relevance of the first ones for astrophysics and, more recently, to the AdS/CFT correspondence for the second ones (for a review see \cite{Aharony:1999ti}). A major part of the research has addressed static, spherical symmetric geometries as computations become significantly more difficult in the case of rotating spacetimes. Apart from the increased number of variables in the problem, the presence of superradiant modes and the intrinsic ambiguity they bring to the definition of positive frequency field modes \cite{Frolov:1989jh,Ottewill:2000qh} contribute to the greater complexity of these cases. Asymptotically AdS spacetimes are usually considered only in the classical regime,  as this is sufficient in the context of the AdS/CFT correspondence, and hence few attempts have been made on studying quantum field theory on these backgrounds.

Many of these issues are greatly simplified if we instead treat lower dimensional spacetimes. (2+1)-dimensional gravity provides a convenient arena to explore several aspects of black hole physics \cite{carlip2003quantum}. The main advantage of this approach is its technical simplicity and, in particular, the fact that many quantities of interest can be obtained in closed form (examples include mode solutions for matter fields and Green's functions for quantum states). It has even been shown that (2+1)-dimensional Einstein gravity is equivalent to a Chern-Simons gauge theory \cite{Achucarro:1987vz,Witten:1988hc}, which provided valuable insights for the quantisations attempts of the theory. Even though Einstein gravity in 2+1 dimensions is a topological theory with no propagating degrees of freedom \cite{carlip2003quantum}, it was possible to find a black hole solution, the Ba\~{n}ados-Teitelboim-Zanelli (BTZ) black hole, when the cosmological constant is negative \cite{Banados:1992wn,Banados:1992gq}. This spacetime is asymptotically AdS and a vast amount of research has been done on it, again partly inspired by the AdS/CFT correspondence (for a review see \cite{Carlip:2005zn}).

Instead of Einstein gravity in (2+1)-dimensions one can also consider a deformation of this theory called Topologically Massive Gravity (TMG), which is obtained by adding a gravitational Chern-Simons term to the Einstein-Hilbert action with a negative cosmological constant \cite{Deser:1982vy,Deser:1981wh}. The resulting theory contains a massive propagating degree of freedom, although at the expense of being a third-order derivative theory. In this sense it is closer in spirit to Einstein gravity in (3+1)-dimensions. One might expect that studying the technically simpler TMG can provide useful insight to some of the challeging problems of the higher dimensional theory. A very important property of this theory is that solutions of Einstein gravity, such as AdS${}_3$ and the BTZ black hole, are also solutions of TMG. Nevertheless, there also exist new kinds of solutions and in this paper we focus on the warped AdS${}_3$ vacuum solutions and warped AdS${}_3$ black hole solutions \cite{Nutku:1993eb,Gurses1994,Moussa:2003fc,Moussa:2008sj,Anninos:2008fx}. Mathematically, warped AdS${}_3$ spacetimes are Hopf fibrations of AdS${}_3$ over AdS${}_2$ where the fibre is the real line and the length of the fibre is `warped' \cite{Bengtsson:2005zj,Anninos:2008qb}. These solutions are thought to be perturbatively stable vacua of TMG in a wide region of the parameter space of the theory, in contrast to the AdS${}_3$ solution \cite{Anninos:2009zi}. Analogously to the BTZ black hole, the warped AdS${}_3$ black hole solutions are obtained as quotients of warped AdS${}_3$ vacuum solutions along Killing directions. In the limit in which the warping of spacetime vanishes one recovers the BTZ black hole as a solution of TMG.

There are several reasons why it is interesting to study classical and quantum matter fields in warped AdS${}_3$ black holes spacetimes. As we shall see in this paper these black holes are rotating and their causal structure resembles asymptotically flat spacetimes in the general case and AdS in the limit of no warping (which corresponds to the BTZ black hole) \cite{Jugeau:2010nq}. We then have at our disposal an example of a (2+1)-dimensional black hole whose asymptotic structure is very similar to the Kerr spacetime and on which we can study both classical stability and field quantisation in a simpler setting. However, these black holes are not asymptotically flat (they are in fact asymptotically warped AdS${}_3$) and this gives us an opportunity to study these matters on the background of a spacetime with a new and unexplored asymptotic structure and also in the context of modified gravity theory. Another particularly novel point is that these rotating black holes do not possess a stationary limit surface (in fact there can be no static observers in the exterior region), but they nonetheless have a speed-of-light surface, which has important consequences in the context of defining thermal quantum states \cite{Frolov:1989jh,Ottewill:2000qh}.

The purpose of this paper is to study in detail a classical massive scalar field on the background of a warped AdS${}_3$ black hole spacetime, with emphasis on classical superradiance and on the stability of the black hole with respect to scalar field perturbations. The existing literature \cite{Oh:2008tc,Kim:2008bf,Chen:2009rf,Compere:2009zj,Blagojevic:2009ek,Chakrabarti:2009ww,Kao:2009fh,Chen:2009hg,Henneaux:2011hv} on the warped AdS${}_3$ solutions is largely motivated by the AdS/CFT correspondence and little attention has been given to these topics. 

Concerning the subject of superradiance, the motivation is two-fold. First, as we alluded above, when attempting to study quantum fields in a rotating black hole spacetime, superradiant field modes have to be treated very carefully. The existence of classical superradiance depends on the boundary conditions that are imposed on the field and on the definition of positive frequency modes, but we shall show that superradiance generally occurs on a warped AdS${}_3$ black hole. This result is valid despite the fact that no stationary limit surface is present. Second, this shows that superradiance is possible in (2+1)-dimensional black hole spacetimes with physically motivated boundary conditions, in contrast to the BTZ black hole for which a particular choice of boundary conditions can be made where superradiance is not present \cite{Ortiz:2011wd}. As a comparison, superradiance is also not inevitable in the (3+1)-dimensional Kerr-AdS black hole \cite{Winstanley:2001nx}, even though it is for Kerr \cite{Frolov:1989jh,Ottewill:2000qh}.

Furthermore, it is fundamental to analyse the classical stability of the black hole to scalar field perturbations before trying to study quantum effects. This is even more important in light of the possibility of superradiant instabilities, which occur in Kerr spacetimes \cite{Press:1972zz,Cardoso:2004nk,Cardoso:2005vk,Dolan:2007mj,Pani:2012vp,Dolan:2012yt,Witek:2012tr}. These instabilities are caused by superradiant modes which are trapped near the event horizon and which over time increase in amplitude. The modes can be trapped either by a mirror surrounding the black hole or by effects due to the mass of the field and in both cases they become unstable bound state modes. In this paper we obtain the bound state modes for a massive scalar field on the background of a warped AdS${}_3$ black hole with and without a mirror. In both cases we conclude that no superradiant instabilities are present. We also obtain the scalar field quasinormal modes, which correspond to the response of the black hole at late times to a perturbation from the scalar field. Quasinormal modes have been extensively studied for a variety of black holes in the last few decades, in many cases in the context of the AdS/CFT correspondence, and are an essential part of perturbation theory of black holes (for recent reviews see \cite{Konoplya:2011qq,Berti:2009kk}). Again, most of the past study on quasinormal and bound state modes have dealt with asymptotically flat and AdS spacetimes and, with the exception of simpler cases like the BTZ black hole \cite{Cardoso:2001hn,Birmingham:2001hc}, it requires numerical methods. For the warped AdS${}_3$ black hole we are able to obtain these modes in closed form and conclude that it is classically stable against these pertubations.

The contents of this paper are as follows. We begin by giving a brief account on TMG in section \ref{sec:tmg} and describe the main features of its warped AdS${}_3$ black hole solutions in section \ref{sec:metric}, with emphasis on their causal structure. We then move on to analyse a massive scalar field on the background of these spacetimes in section \ref{sec:KG}. In section \ref{sec:fieldeq} we solve the Klein-Gordon field equation and then discuss the existence of classical superradiance in section \ref{sec:superradiance}. Having dealt with this issue we proceed to obtain the quasinormal and bound state modes for the scalar field in section \ref{sec:qnm} and comment on the classical stability of the black hole against the scalar perturbation. In section \ref{sec:mirror} we verify that the previous analysis is not changed upon the introduction of a mirror in the exterior region of the spacetime. Finally, in section \ref{sec:conclusions} we present the conclusions.

%%%%%%%%%%%%%%%%%%%%%%%%%%%%%%%%%%%%%%%%%%%%%%%%%%%%%%%%%%%%%%%%%%

%%% SPACELIKE STRETCHED BLACK HOLES

\section{Warped A\lowercase{d}S black holes}

In this section we give a short description of Topologically Massive Gravity and review the basic features of the warped AdS${}_3$ black hole solutions, including their causal structure.

%% TOPOLOGICALLY MASSIVE GRAVITY

\subsection{Topologically Massive Gravity}
\label{sec:tmg}

The action of Topologically Massive Gravity (TMG) in 2+1 spacetime dimensions is obtained by adding a gravitational Chern-Simons term to the Einstein-Hilbert action with a negative cosmological constant \cite{Deser:1982vy,Deser:1981wh}
\begin{equation}
S = S_{\text{E-H}} + S_{\text{C-S}} \, ,
\end{equation}
with
\begin{align}
S_{\text{E-H}} &= \frac{1}{16\pi G} \int d^3 x \sqrt{-g} \left( R + \frac{2}{\ell^2} \right) \, , \\
S_{\text{C-S}} &= \frac{\ell}{96\pi G \nu} \int d^3 x \sqrt{-g} \, \epsilon^{\lambda\mu\nu} \, \Gamma^{\rho}_{\lambda\sigma} \left( \partial_{\mu} \Gamma^{\sigma}_{\rho\nu} + \frac{2}{3} \Gamma^{\sigma}_{\mu\tau} \Gamma^{\tau}_{\nu\rho} \right) \, .
\end{align}
$G$ is Newton's gravitational constant, $\nu$ is a dimensionless coupling, $g$ is the determinant of the metric, $R$ is the Ricci scalar, $\ell > 0$ is the cosmological length (related to the cosmological constant $\Lambda$ by $\Lambda = - 1/\ell^2$), $\Gamma^{\rho}_{\lambda\sigma}$ are the Christoffel symbols and $\epsilon^{\lambda\mu\nu}$ is the Levi-Civita tensor in 3 dimensions. We use the $(-,+,+)$ signature and from now on units such that $\hbar = c = G = 1$.

Contrary to Einstein gravity in three dimensions \cite{carlip2003quantum}, TMG has a massive propagating degree of freedom, while at the same retaining the BTZ black hole \cite{Banados:1992wn,Banados:1992gq} as a solution. Nevertheless, there also exist solutions of TMG which are not solutions of Einstein gravity, including the warped AdS${}_3$ vacuum solutions and warped AdS${}_3$ black hole solutions \cite{Nutku:1993eb,Gurses1994,Moussa:2003fc,Moussa:2008sj,Anninos:2008fx,Bengtsson:2005zj,Anninos:2008qb}. Similarly to the BTZ solution, the latter are obtained from the former by global identifications. In the next section we describe in more detail one type of these warped AdS${}_3$ black hole solutions, the spacelike stretched black hole.

%% METRIC

\subsection{Geometry and causal structure of the spacelike stretched black hole}
\label{sec:metric}

The spacelike stretched black hole metric in coordinates $(t,r,\theta)$ is \cite{Anninos:2008fx}
\begin{equation}
ds^2 = dt^2 + \frac{\ell^2 dr^2}{4 R^2(r) N^2(r)} + 2 R^2(r) N^{\theta}(r) dt d\theta + R^2(r) d\theta^2 \, ,
\label{eq:metricbh}
\end{equation}
with $r \in (0,\infty)$, $t \in (-\infty,\infty)$, $(t,r,\theta) \sim (t,r,\theta + 2\pi)$ and
\begin{align}
R^2(r) &= \frac{r}{4} \left[ 3(\nu^2-1)r + (\nu^2+3)(r_+ + r_-) - 4\nu \sqrt{r_+ r_-(\nu^2+3)} \right] \, , \\
N^2(r) &= \frac{(\nu^2+3)(r-r_+)(r-r_-)}{4R(r)^2} \, , \\
N^{\theta}(r) &= \frac{2\nu r - \sqrt{r_+ r_- (\nu^2+3)}}{2 R(r)^2} \, .
\end{align}

There are outer and inner event horizons at $r = r_+$ and $r = r_-$, respectively, and a curvature singularity located at $r = \bar{r}_0 \equiv \max\{ 0, r_0 \}$, with
\begin{equation}
r_0 = \frac{4\nu \sqrt{r_+ r_- (\nu^2+3)} - (\nu^2+3)(r_++r_-)}{3(\nu^2-1)} \, ,
\end{equation}
which satisfies $R(\bar{r}_0) = 0$. These quantities obey the inequalities $0 \leq \bar{r}_0 \leq r_- \leq r_+$ \cite{Anninos:2008fx,Jugeau:2010nq}. The dimensionless coupling $\nu$ is in this context called the warp factor and for the spacelike stretched black hole its domain is $\nu > 1$. In the limit $\nu \to 1$ the above metric reduces to the metric of the BTZ black hole in a rotating frame.

Similarly to other rotating black hole solutions, the vector fields $\partial_t$ and $\partial_{\theta}$ are Killing vectors fields in the coordinate system used above. However, it is immediately clear that $\partial_t$ is spacelike \emph{everywhere} in the spacetime, even though surfaces of constant $t$ are still spacelike outside the event horizon. Consequently, there is no stationary limit surface and, more importantly, no observers following orbits of $\partial_t$ (the usual `static observers' in other spacetimes) anywhere in the exterior region.

Nonetheless, one can still consider observers at a given radius $r$ following orbits of the vector field $\xi(r) = \partial_t + \Omega(r) \, \partial_{\theta}$, which are timelike as long as
\begin{equation}
\Omega_-(r) < \Omega(r) < \Omega_+(r) \, ,
\label{eq:conditionalmoststationary}
\end{equation}
with
\begin{equation}
\Omega_{\pm}(r) = - \frac{2}{2\nu r - \sqrt{r_+ r_- (\nu^2+3)} \pm \sqrt{(r-r_+)(r-r_-)(\nu^2+3)}} \, .
\end{equation}
$\Omega(r)$ is negative for all $r > r_+$, approaches zero as $r \to +\infty$ and tends to
\begin{equation}
\Omega_{\mathcal{H}} = - \frac{2}{2\nu r_+ - \sqrt{r_+ r_- (\nu^2+3)}}
\label{eq:omegaH}
\end{equation}
as $r \to r_+$. In view of these observations, we can regard $\Omega_{\mathcal{H}}$ as the angular velocity of the horizon with respect to stationary observers close to infinity.

Even though there is no stationary limit surface, there is still a speed-of-light surface, beyond which an observer cannot co-rotate with the event horizon. Given the information above it is easy to check that the vector field $\chi = \partial_t + \Omega_{\mathcal{H}} \, \partial_{\theta}$ is the Killing vector field which generates the horizon. $\chi$ is null at the horizon and at
\begin{equation}
r = r_{\mathcal{C}} = \frac{4\nu^2 r_+ - (\nu^2+3) r_-}{3(\nu^2-1)} \, ,
\end{equation}
which is the location of the speed-of-light surface. This surface is important in the context of quantum field theory, in particular in defining thermal states \cite{Frolov:1989jh,Ottewill:2000qh}.

As noted before the spacetime is \emph{not} asymptotically AdS${}_3$. The asymptotic form of the metric \eqref{eq:metricbh} at $r \to +\infty$ is
\begin{equation}
ds^2 = dt^2 + \frac{\ell^2 dr^2}{(\nu^2+3)r^2} + 2\nu r \, dt d\theta + \frac{3(\nu^2-1)r^2}{4} d\theta^2 \, ,
\label{eq:asymptoticmetric}
\end{equation}
which is locally the metric of the spacelike stretched AdS${}_3$ spacetime (in \eqref{eq:asymptoticmetric} $\theta$ is periodic whereas in the spacelike stretched AdS${}_3$ it is unwrapped) \cite{Anninos:2008fx,Anninos:2009zi}. In this paper we adopt the usual abuse of language and say that the spacelike stretched black hole solutions are asymptotic to spacelike stretched AdS${}_3$.

\begin{figure}[t!]
\begin{center}
\subfigure[$\; r_0 < r_- < r_+$]{
\includegraphics[scale=0.55]{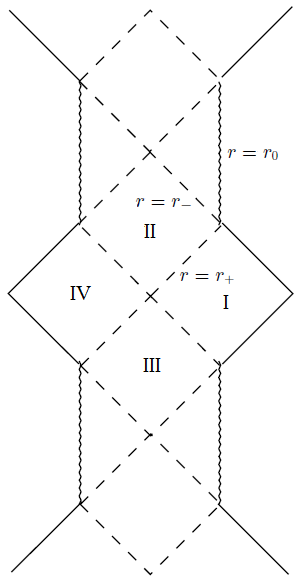}}
\subfigure[$\; r_0 = r_- < r_+$]{
\includegraphics[scale=0.5]{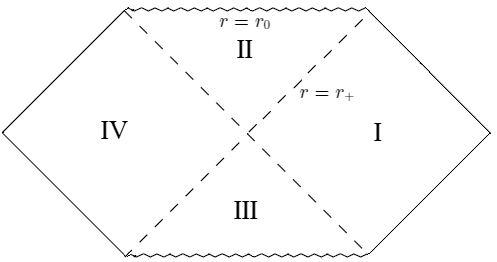}}
\subfigure[$\; r_0 < r_- = r_+$]{
\includegraphics[scale=0.52]{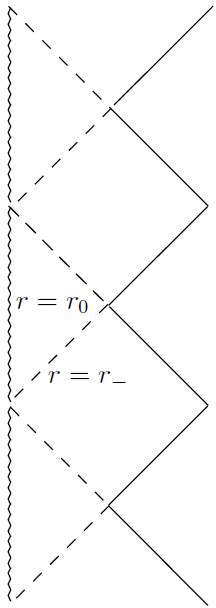}}
\caption{\label{fig:CPdiagrams} Carter-Penrose diagrams of the spacelike stretched black hole spacetime for different values of $r_0$, $r_-$ and $r_+$ (adapted from \cite{Jugeau:2010nq}).}
\end{center}
\end{figure}

From the Carter-Penrose diagrams, shown in Fig.~\ref{fig:CPdiagrams}, we see that the causal structure is very similar to that of asymptotically flat spacetimes in 3+1 dimensions. Indeed, the diagrams for the cases $r_0 < r_- < r_+$ and $r_0 < r_- = r_+$ are exactly the same as those for the standard and extreme Reissner-Nordstr\"{o}m black holes, while the one for the case $r_0 = r_- < r_+$ is identical to that for the Kruskal spacetime. For this reason one may expect the behaviour of matter fields on these spacetimes to be qualitatively similar to that on the asymptotically flat ones. In the next section we proceed to study classical scalar fields on spacelike stretched black holes.

%%%%%%%%%%%%%%%%%%%%%%%%%%%%%%%%%%%%%%%%%%%%%%%%%%%%%%%%%%%%%%%%%%

%%% MASSIVE SCALAR FIELD FIELD AND CLASSICAL SUPERRADIANCE

\section{Massive Scalar field and classical superradiance}
\label{sec:KG}

In this section we start by obtaining the solutions of the classical Klein-Gordon equation for a real massive scalar field on the background of the spacelike stretched black hole, both in closed form and asymptotic approximations near the event horizon and infinity. We then construct sets of basis modes and use them to discuss the existence of classical superradiance.

%% FIELD EQUATION

\subsection{Klein-Gordon Field equation}
\label{sec:fieldeq}

We consider a real massive scalar field $\Phi$ on a spacelike stretched black hole. The field obeys the Klein-Gordon equation
\begin{equation}
\left(\nabla^2 - m_0^2 - \xi R \right) \Phi = 0 \, ,
\label{eq:fieldequation1}
\end{equation}
where $m_0$ is the mass of the field, $R$ is the Ricci scalar and $\xi$ is the curvature coupling parameter. For our spacetime $R =  - (\nu^2+3)+(\nu^2-3)/\ell^2$, which is a constant, so we can rewrite \eqref{eq:fieldequation1} as
\begin{equation}
\left(\nabla^2 - m^2 \right) \Phi = 0 \, ,
\label{eq:fieldequation2}
\end{equation}
where $m^2 \equiv m_0^2 + \xi R$ is the `effective squared mass' of the scalar field.

Since $\partial_t$ and $\partial_{\theta}$ are Killing vector fields of the spacetime, we consider solutions of \eqref{eq:fieldequation2} of the form
\begin{equation}
\Phi(t,r,\theta) = e^{-i\omega t + i k \theta} \, \phi(r) \, ,
\label{eq:fieldansatz}
\end{equation}
where the frequency $\omega$ is continuous and the angular momentum number $k$ is an integer.

Using \eqref{eq:fieldansatz} one gets the radial equation
\begin{equation}
\frac{4}{\ell^2} R^2 N^2 \frac{d}{dr} \left( R^2 N^2 \frac{d\phi}{dr} \right) + \left[ R^2 (\omega + k N^{\theta})^2 - N^2 (k^2 + m^2 R^2) \right] \phi = 0 \, .
\label{eq:radialeq}
\end{equation} 
By performing the rescalings $r \to r \ell$, $t \to t \ell$, $m \to m/\ell$ and $\omega \to \omega/\ell$, one can set $\ell=1$, as is assumed from now on.

It is possible to obtain an exact solution to the radial equation. If we introduce the coordinate
\begin{equation}
z = \frac{r-r_+}{r-r_-} \, 
\label{eq:definitionz}
\end{equation}
the general solution can be written as
\begin{equation}
\phi_{\omega k} (z) = A_{\omega k} \, z^{\alpha} (1-z)^{\beta} F(a,b,c;z) + B_{\omega k} \, z^{\alpha^*} (1-z)^{\beta^*} F^*(a,b,c;z) \, ,
\label{eq:exactsolution}
\end{equation}
where $A_{\omega k}$ and $B_{\omega k}$ are constants and the parameters of the hypergeometric functions are given by
\begin{align}
a = \alpha + \beta + \gamma \, , \qquad
b = \alpha + \beta - \gamma \, , \qquad
c = 2\alpha + 1 \, ,
\label{eq:defabc}
\end{align}
where
\begin{align}
\alpha &= -i \tilde{\omega}_+ \frac{2\nu r_+ - \sqrt{r_+ r_-(\nu^2+3)}}{(\nu^2+3)(r_+ - r_-)} \, , \\
\beta &= \frac{1}{2} - \hat{\varpi} \frac{\sqrt{3(\nu^2-1)}}{\nu^2+3} \, , \\ 
\gamma & = -i \tilde{\omega}_- \frac{2\nu r_- - \sqrt{r_+ r_-(\nu^2+3)}}{(\nu^2+3)(r_+ - r_-)} \, ,
\end{align}
and
\begin{align}
\tilde{\omega}_{\pm} \equiv \omega + k N^{\theta}(r_{\pm}) \, , \qquad
\hat{\varpi} \equiv \sqrt{\frac{(\nu^2+3)^2}{12(\nu^2-1)} \left( 1 + \frac{4m^2}{\nu^2+3} \right) - \omega^2} \, .
\end{align}

This exact solution will be useful for the stability analysis below. To discuss the existence of classical superradiance it will be sufficient to consider approximations near the horizon and infinity, as is done in the next subsection.

%% BASIS MODES AND SUPERRADIANCE

\subsection{Basis modes and classical superradiance}
\label{sec:superradiance}

In this subsection we construct sets of basis modes with which a general solution of the field equation can be expressed at any point of the spacetime. Although it is possible to write down exact solutions in terms of hypergeometric functions as above, a description in terms of these sets of basis modes is very useful in many contexts, in particular when discussing the phenomenon of superradiance.

\begin{figure}[t!]
\centering
\includegraphics[width=0.6\textwidth]{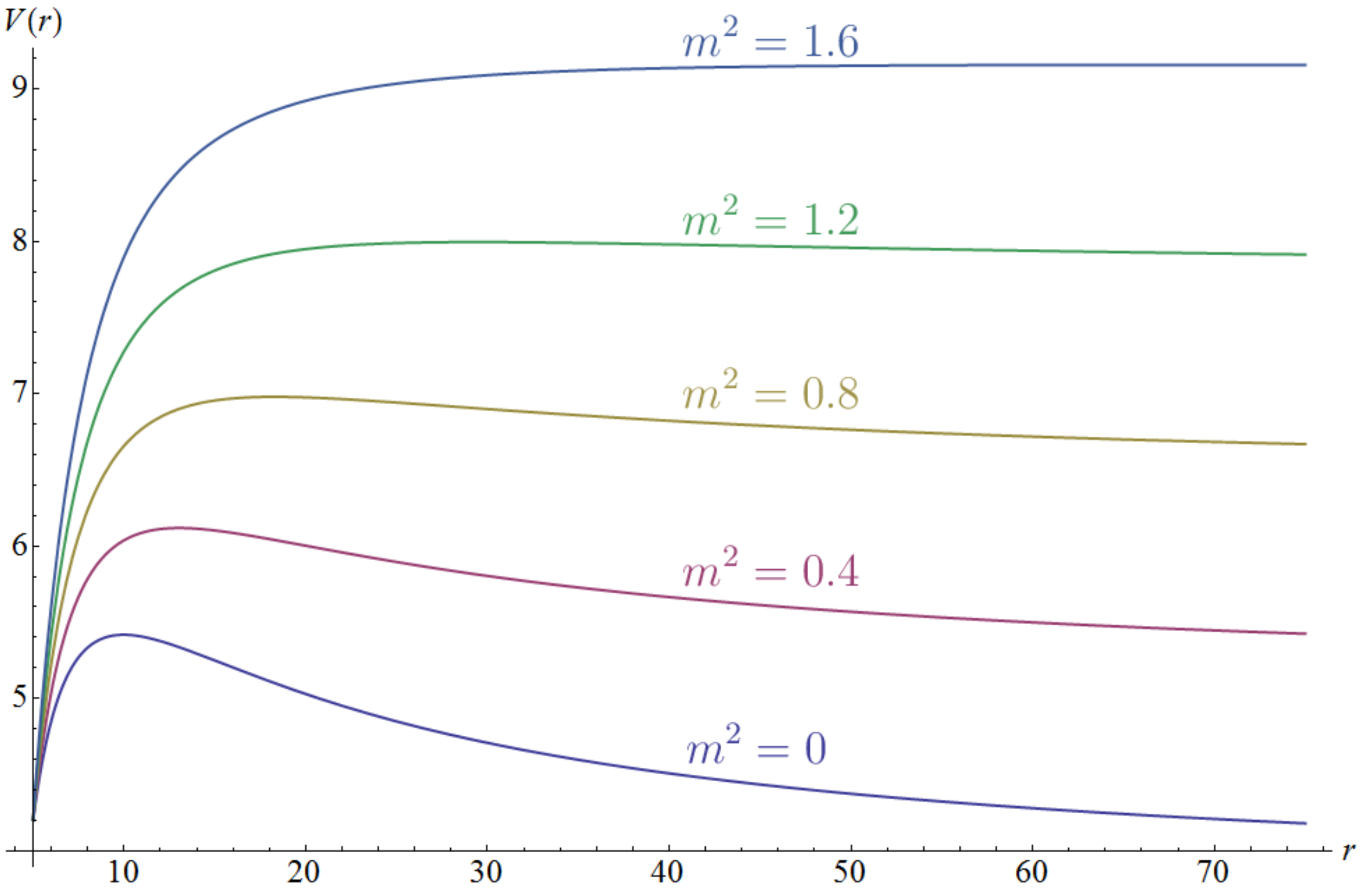}
\caption{Effective potential $V_{\omega k}(r)$ for selected values of $m^2$ with $r_+=5$, $r_-=2.5$, $\nu=1.2$, $\omega = 5$ and $k=-1$. For smaller values of $m^2$ (or larger values of $\omega$) the potential has a local maximum near the horizon, around which a potential barrier stands. As one considers fields with larger $m^2$ (or smaller $\omega$) the potential barrier eventually disappears.}
\label{fig:potentialmass}
\end{figure}

First, it will be important to obtain the effective potential seen by the scalar field. To do that we define the tortoise coordinate $r_*$ by the equation
\begin{equation}
\frac{dr_*}{dr} = \frac{1}{2 R N^2} \, ,
\label{eq:tortoiseeq}
\end{equation}
whose domain is $(-\infty,\infty)$ for $\nu>1$, and introduce the new radial function $\varphi (r)$ by
\begin{equation}
\phi_{\omega k}(r) \equiv R(r)^{-1/2} \, \varphi_{\omega k} (r) \, .
\label{eq:varphi}
\end{equation}
The radial equation \eqref{eq:radialeq} can then be written as
\begin{equation}
\left( \frac{d^2}{dr_*^2} + (\omega^2 - V_{\omega k}(r)) \right) \varphi_{\omega k}(r) = 0 \, ,
\label{eq:effectiveeq}
\end{equation}
with:
\begin{equation}
V_{\omega k} \equiv \omega^2 - (\omega + k N^{\theta})^2 + 2 N^3 \left( R N \frac{d^2 R}{dr^2} + \frac{1}{2} N \left(\frac{d R}{dr}\right)^2 + 2 R \frac{d R}{dr} \frac{d N}{dr} \right) + N^2 \left( m^2 + \frac{k^2}{R^2} \right) \, .
\label{eq:effectivepotential}
\end{equation}
We may hence regard $V_{\omega k}(r)$ as the effective potential experienced by the scalar field of effective squared mass $m^2$, frequency $\omega$ and angular momentum number $k$. Similarly to what happens in the Kerr spacetime \cite{Simone:1991wn}, $V_{\omega k}(r)$ depends on the frequency $\omega$ of the scalar field (when $k \neq 0$). Also, $V_{\omega k}(r) \to +\infty$ as $r \to +\infty$ and $\nu \to 1$, as expected for the BTZ black hole. Note that in this case the domain of the tortoise coordinate $r_*$ becomes $(-\infty,\hat{r}_*)$, where $\hat{r}_*$ is a finite value. Fig.~\ref{fig:potentialmass} shows the form of $V_{\omega k}(r)$ for selected values of $m^2$.

In the near-horizon limit we have
\begin{equation}
V_{\omega k} - \omega^2 \simeq - \tilde{\omega}^2 \, ,
\end{equation}
where $\tilde{\omega} \equiv \omega + k N^{\theta}(r_+) = \omega - k \Omega_{\mathcal{H}}$. Thus the solution near the horizon is of the form
\begin{equation}
\varphi_{\omega k}(r_*) = A_{\omega k} \, e^{i \tilde{\omega} r_*} + B_{\omega k} \, e^{- i \tilde{\omega} r_*} \, .
\end{equation}
Modes of the form $e^{i \tilde{\omega} r_*}$ are outcoming from the past event horizon, while modes of the form $e^{-i \tilde{\omega} r_*}$ are ingoing to the future event horizon.

At infinity we have
\begin{align}
V_{\omega k} \to \omega_m^2 \, ,
\end{align}
where
\begin{equation}
\omega_m \equiv \frac{1}{2} \frac{\nu^2+3}{\sqrt{3(\nu^2-1)}} \sqrt{ 1 + \frac{4m^2}{\nu^2+3} } \, .
\label{eq:omegam}
\end{equation}
This behaviour contrasts with that in asymptotically flat spacetimes such as Kerr, where the effective potential tends to $m^2$ at infinity \cite{Simone:1991wn} and with asymptotically AdS spacetimes such as the BTZ or Kerr-AdS, where the effective potential grows without bound at infinity \cite{Winstanley:2001nx}. We assume that the asymptotic value of $V_{\omega k}$ at infinity, $\omega_m^2$, is non-negative and, by \eqref{eq:omegam}, this implies that $m^2$ may be negative provided it satisfies $m^2 \geq -\frac{\nu^2+3}{4}$. As a consistency check, in the BTZ limit $\nu \to 1$ this inequality reduces to the Breitenlohner-Freedman bound for AdS${}_3$ spacetimes $m^2 \geq -1$ \cite{Breitenlohner:1982bm}.

In the case $|\omega| > \omega_m$, near infinity the solution is of the form
\begin{equation}
\varphi_{\omega k}(r_*) = C_{\omega k} \, e^{i \hat{\omega} r_*} + D_{\omega k} \, e^{- i \hat{\omega} r_*} \, ,
\end{equation}
where
\begin{equation}
\hat{\omega} \equiv \begin{cases}
\sqrt{\omega^2-\omega_m^2} \, , & \omega > \omega_m \geq 0 \, ; \\
-\sqrt{\omega^2-\omega_m^2} \, , & \omega < -\omega_m \leq 0 \, .
\end{cases}
\label{eq:hatomegadef}
\end{equation}
When $\hat{\omega} > 0$, modes of the form $e^{i \hat{\omega} r_*}$ correspond to outgoing flux at infinity, while the modes of the form $e^{-i \hat{\omega} r_*}$ correspond to incoming flux at infinity and vice-versa when $\hat{\omega} < 0$. One way to see this is by calculating the radial flux $j^r$ of the field mode $\phi_{\omega k}$ at infinity
\begin{equation}
j^r = - i \, g^{rr} \left( \phi_{\omega k}^* \, \partial_r \phi_{\omega k} - \phi_{\omega k} \, \partial_r \phi_{\omega k}^* \right) \, .
\end{equation}
The result turns out to be
\begin{equation}
j^r = 4 \hat{\omega} \left( |C_{\omega k}|^2 - |D_{\omega k}|^2 \right) \, , \qquad r \to +\infty
\end{equation}
Since a positive (negative) radial flux at infinity corresponds to outgoing (incoming) flux, the interpretation above follows.

Finally, if $|\omega| < \omega_m$, the solutions near infinity are
\begin{equation}
\varphi_{\omega k}(r_*) = E_{\omega k} \, e^{\hat{\varpi} r_*} + F_{\omega k} \, e^{-\hat{\varpi} r_*} \, ,
\end{equation}
where
\begin{equation}
\hat{\varpi} \equiv \begin{cases}
\sqrt{\omega_m^2 - \omega^2} \, , & 0 < \omega < \omega_m \, ; \\
-\sqrt{\omega_m^2 - \omega^2} \, , & -\omega_m < \omega < 0 \, .
\end{cases}
\label{eq:hatvarpidef}
\end{equation}
To exclude the solution that diverges exponentially at infinity, we impose that $E_{\omega k} = 0$ when $0 < \omega < \omega_m$ and  $F_{\omega k} = 0$ when $-\omega_m < \omega < 0$.

Note that so far we have not made any choice of `positive frequency', for instance by taking $\omega > 0$. We will return to this point at the end of this subsection.

\begin{figure}[t!]
\begin{center}
\subfigure[\; `In' modes]{
\includegraphics[scale=0.4]{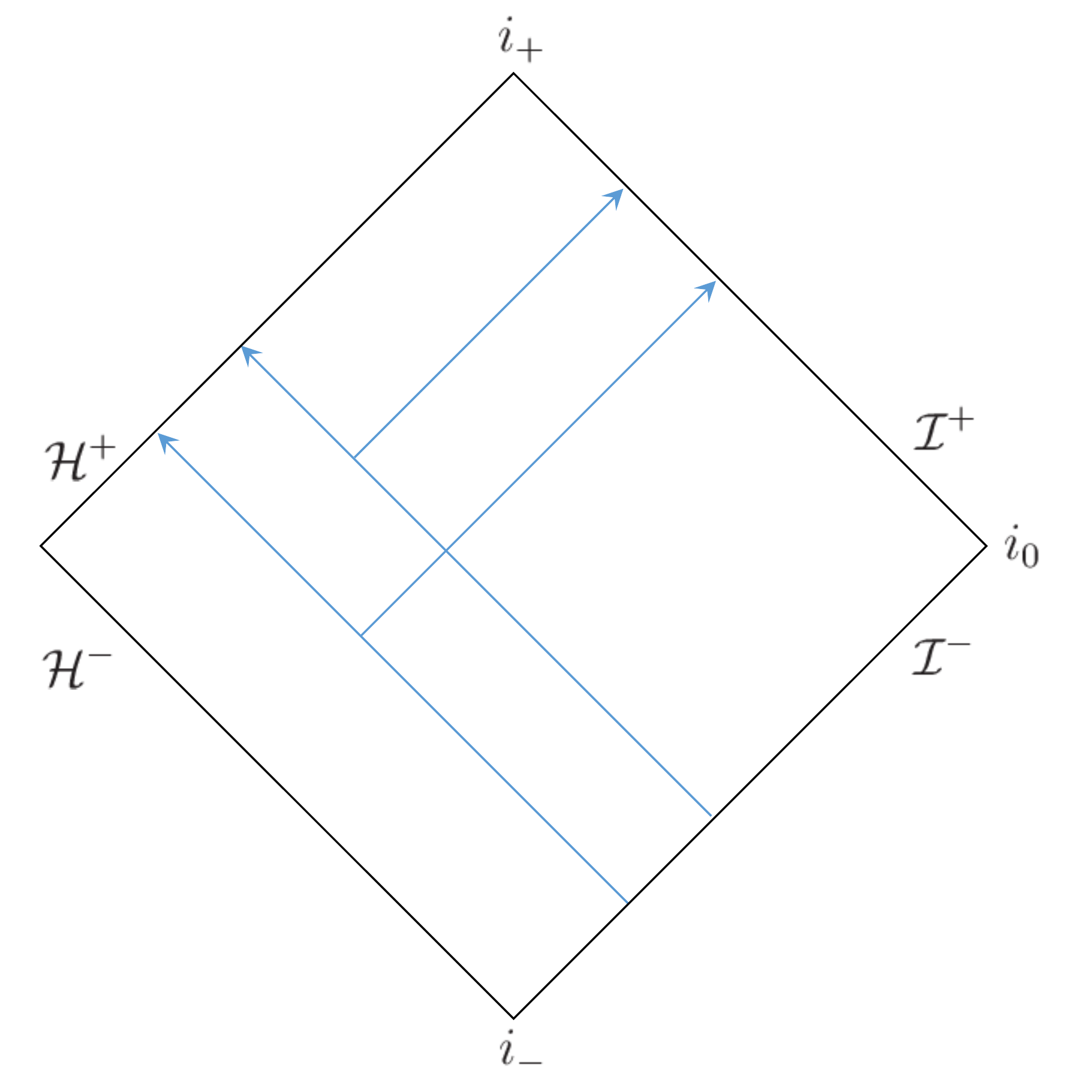}}
\subfigure[\; `Up' modes]{
\includegraphics[scale=0.4]{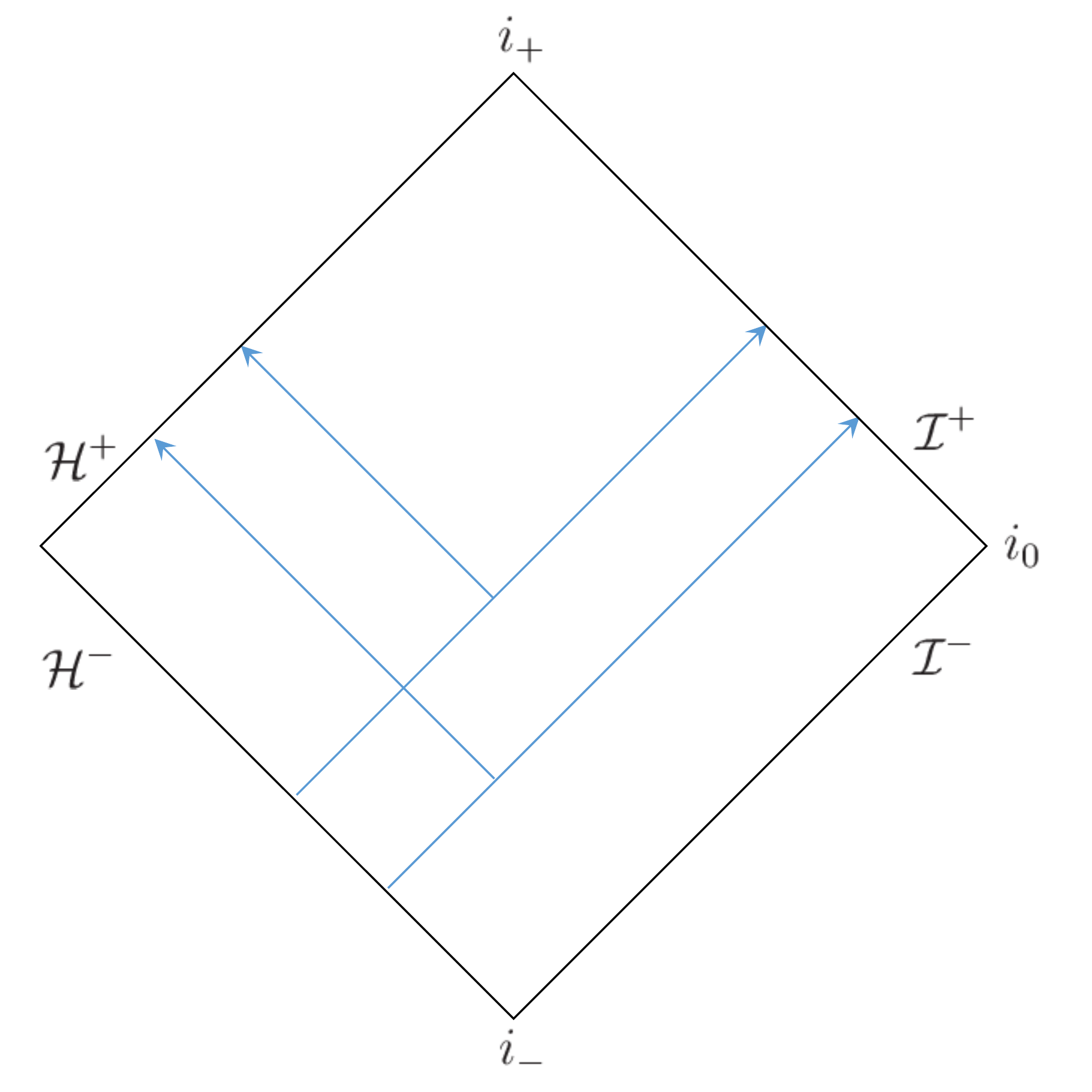}}
\caption{\label{fig:inupmodes} `In' and `up' modes in the exterior region of the spacetime.}
\end{center}
\end{figure}

We are now in possession of the tools needed to construct a set of basis modes. Two particular basis modes will be of particular importance in the following, the `in' and `up' modes, which are specified by the boundary conditions they obey at the event horizon and at infinity. These modes are defined in analogy with the Kerr spacetime \cite{Ford:1975tp,Frolov:1989jh}. For $|\omega| > \omega_m$, the `in' modes satisfy
\begin{equation}
\varphi^{\text{in}}_{\omega k}(r_*) = \begin{cases}
B^{\text{in}}_{\omega k} \, e^{- i \tilde{\omega} r_*} \, , & r_* \to -\infty \\
e^{- i \hat{\omega} r_*} + C^{\text{in}}_{\omega k} \, e^{i \hat{\omega} r_*} \, , & r_* \to +\infty
\end{cases}
\end{equation}
whereas the `up' modes satisfy
\begin{equation}
\varphi^{\text{up}}_{\omega k}(r_*) = \begin{cases}
e^{i \tilde{\omega} r_*} + B^{\text{up}}_{\omega k} \, e^{- i \tilde{\omega} r_*} \, , & r_* \to -\infty \\
C^{\text{up}}_{\omega k} \, e^{i \hat{\omega} r_*} \, , & r_* \to +\infty
\end{cases}
\end{equation}
The `in' modes correspond to flux coming from infinity which is partially reflected back to infinity and partially absorbed by the black hole. The `up' modes correspond to flux coming from the black hole which is partially reflected back to the black hole and partially sent to infinity. This is represented in Fig.~\ref{fig:inupmodes}. Note that if $\omega_m > 0$ there are in addition bound state modes for which $|\omega| < \omega_m$ (we return to these at the end of this subsection).

These two solutions are linearly independent and suffice to describe any general solution (with $|\omega| > \omega_m$) in any point of the exterior region of the spacetime. Given any two linearly independent solutions $\varphi_1(r_*)$ and $\varphi_2(r_*)$, their Wroskian is constant, i.e.
\begin{equation}
W(\varphi_1, \varphi_2) \equiv \varphi_1 {\varphi^*_2}' - \varphi_1' \varphi^*_2 = \text{constant}
\end{equation}
By applying this to the `in' and `up' modes, we obtain
\begin{equation}
\tilde{\omega} \, |B^{\text{in}}_{\omega k}|^2 = \hat{\omega} \left( 1 - |C^{\text{in}}_{\omega k}|^2 \right) \, , \qquad
\tilde{\omega} \left( 1 - |B^{\text{up}}_{\omega k}|^2 \right) = \hat{\omega} \, |C^{\text{up}}_{\omega k}|^2 \, .
\end{equation}
An incident `in' mode from infinity is hence reflected back to infinity with a greater amplitude, $|C^{\text{in}}_{\omega k}|>1$, if and only if $\tilde{\omega} \hat{\omega} < 0$. Similarly, an incident `up' mode from the horizon is reflected back to the horizon with an increased amplitude, $|B^{\text{up}}_{\omega k}|>1$, if and only if $\tilde{\omega} \hat{\omega} < 0$. This phenomenon is known as \emph{superradiance}.

At this point we need to discuss the notion of positive frequency, i.e. we need to decide the location of a locally non-rotating observer (LNRO) with respect to whom only positive frequency modes are observed. We adopt the terminology of \cite{Frolov:1989jh} concerning `near-horizon' and `distant' observers. 

For the `in' modes it is conventional to have positive frequency as measured by a LNRO at infinity (the `distant observer' viewpoint), meaning $\omega > 0$. If $\omega_m > 0$ we must additionally have $\omega > \omega_m$ for the `in' mode to exist, so that the positive frequency condition altogether is $\omega > \omega_m$. There is thus the possibility that $\tilde{\omega} < 0$ if $\omega_m < \omega < k \Omega_{\mathcal{H}}$. Therefore, `in' superradiant modes can exist with this choice of positive frequency. 

For the `up' modes it is conventional to have positive frequency as measured by a LNRO close to the horizon (the `near-horizon observer' viewpoint), meaning $\tilde{\omega} > 0$. Thus, an `up' mode with $\omega < - \omega_m$ has $\hat{\omega} < 0$ and is an `up' superradiant mode with this choice of positive frequency.

However, if one swaps the viewpoints it is easy to check that one can still have superradiant modes in both cases. One thus concludes that whatever the choice of positive frequency there is always the possibility to have superradiant field modes in the spacelike stretched black hole. This is in agreement with the expectation that the behaviour of the field modes should be similar to the Kerr spacetime case, given the similar causal structure and boundary conditions that we imposed. It is also interesting to note that the situation is significantly different for the Kerr-AdS spacetime, where classical superradiance is not inevitable \cite{Winstanley:2001nx}.

Finally, suppose that $\omega_m > 0$ and consider the bound state modes
\begin{equation}
\varphi^{\text{bs}}_{\omega k}(r_*) = \begin{cases}
A^{\text{bs}}_{\omega k} \, e^{i \tilde{\omega} r_*} + B^{\text{bs}}_{\omega k} \, e^{- i \tilde{\omega} r_*} \, , & r_* \to -\infty \\
e^{- \hat{\varpi} r_*} \, , & r_* \to +\infty
\end{cases}
\end{equation}
where $0 < \omega < \omega_m$. The Wronskian relation reads $|A^{\text{bs}}_{\omega k}| = |B^{\text{bs}}_{\omega k}|$, implying that all flux coming from the horizon is reflected back. Consequently, in this frequency range there are no superradiant modes. This is similar to the situation with the BTZ black hole when reflective boundary conditions are imposed \cite{Ortiz:2011wd}.

%%%%%%%%%%%%%%%%%%%%%%%%%%%%%%%%%%%%%%%%%%%%%%%%%%%%%%%%%%%%%%%%%%

%%% CLASSICAL STABILITY AGAINST SCALAR PERTURBATIONS

\section{Classical stability against scalar perturbations}
\label{sec:stability}

In this section we find the quasinormal and bound state scalar field modes and use the results to comment on the classical stability of the black holes solutions to massive scalar perturbations. We then analyse the case in which a mirror-like boundary is added to the exterior region of the spacetime and show that the previous stability conclusions are unchanged.

%% QUASINORMAL MODES

\subsection{Quasinormal and bound state modes}
\label{sec:qnm}

Suppose that a spacelike stretched black hole is perturbed by a massive scalar field propagating in the spacetime. Once the black hole is perturbed it responds by releasing gravitational and scalar waves in the form of characteristic quasinormal modes of discrete complex frequencies (for recent reviews on quasinormal modes see \cite{Konoplya:2011qq,Berti:2009kk}). For a stable black hole the quasinormal modes are exponentially decaying in time; conversely, if any of the modes are increasing in time, the black hole is unstable. Moreover, as we saw in the previous section, there can be superradiant modes in this spacetime. If any of these superradiant modes are localised near the event horizon in the form of bound states modes (possibly due to a potential well in the effective potential felt by the scalar field) the repeated amplitude increases due to reflections on the walls of the potential well lead to the so-called superradiant instabilities \cite{Press:1972zz,Cardoso:2004nk,Cardoso:2005vk,Dolan:2007mj,Pani:2012vp,Dolan:2012yt,Witek:2012tr}.

The quasinormal and bound state modes are defined by appropriate boundary conditions at the horizon and at infinity. Since we are studying the system at a classical level there must be no flux from the horizon and thus we impose that only ingoing modes are present. Furthermore, we want no perturbations coming in from infinity and hence we require that the quasinormal modes obey outgoing boundary conditions at infinity. As for the bound state modes, since they are localised in the vicinity of the black hole, we impose that they decrease exponentially at infinity.

These boundary conditions on field modes with a $e^{-i \omega t}$ time dependence restricts the allowed frequencies $\omega$ to a discrete set of complex values. The real part of $\omega$ represents the physical frequency of the oscillation whereas the imaginary part gives the decay (or growth) in time of the mode. This occurs because the field can escape to the black hole or to infinity. We are then faced with an eigenvalue problem in which the quasinormal or the bound state modes are the eigenmodes. By obtaining the eigenfrequencies we can infer the stability of a given mode by the sign of the imaginary part: if the imaginary part is negative then the mode decays in time and does not create an instability.

We now proceed to calculate the complex frequencies of the quasinormal and bound state modes. Contrary to most higher dimensional black hole spacetimes, in our case we do not have to resort to numerical methods since we have an analytical expression for the field modes \eqref{eq:exactsolution} to which we can apply the above boundary conditions. By imposing the ingoing boundary condition at the horizon to the solution we are left with
\begin{equation}
\phi_{\omega k}(z) = A_{\omega k} \, z^{\alpha} (1-z)^{\beta} F(a,b,c;z) \, .
\end{equation}
In order to impose the boundary condition at infinity one uses one of the connection formulas relating hypergeometric functions at different regular singularities \cite{NISTbook}, resulting in
\begin{align}
\phi_{\omega k}(z) &= A_{\omega k} \, \Gamma(c) \, z^{\alpha} \Bigg[ (1-z)^{\beta} \frac{\Gamma(c-a-b)}{\Gamma(c-a) \Gamma(c-b)} F(a,b,a+b-c+1;1-z) \notag \\
&\qquad\qquad\qquad\; + (1-z)^{\beta^*} \, \frac{\Gamma(a+b-c)}{\Gamma(a) \Gamma(b)} F(c-a,c-b,c-a-b+1;1-z) \Bigg] \, .
\label{eq:QNandBSmodes}
\end{align}
Note that at infinity
\begin{equation}
(1-z)^{\beta} \sim r^{-1/2} e^{i \hat{\omega} r_*} = r^{-1/2} e^{\hat{\varpi} r_*} \, , \qquad
(1-z)^{\beta^*} \sim r^{-1/2} e^{-i \hat{\omega} r_*} = r^{-1/2} e^{-\hat{\varpi} r_*} \, ,
\label{eq:behaviouratinfinity}
\end{equation}
where $\hat\omega$ and $\hat{\varpi}$ were defined in \eqref{eq:hatomegadef} and \eqref{eq:hatvarpidef}, respectively. The frequency $\omega$ is complex and we write it as $\omega = \omega_{\text{R}} + i \omega_{\text{I}}$. One can use the $(t,\theta) \to (-t,-\theta)$ symmetry to only consider solutions with $\omega_{\text{R}} \geq 0$ and thus $\text{Re}[\hat\omega] \geq 0$, $\text{Re}[\hat\varpi] \geq 0$. 

As we described above a quasinormal mode is characterised for being outgoing at infinity, which implies that the $(1-z)^{\beta^*}$ term in \eqref{eq:QNandBSmodes} must vanish. This happens when
\begin{equation}
a = -n \quad \text{or} \quad b = - n \, ,
\label{eq:anb}
\end{equation}
where $n \in \mathbb{N}_0$ is the overtone number.

On the other hand a bound state mode must exponentially decrease at spatial infinity and hence the $(1-z)^{\beta}$ term in \eqref{eq:QNandBSmodes} must vanish. This happens when
\begin{equation}
c - a = -n \quad \text{or} \quad c - b = - n \, .
\label{eq:cacbn}
\end{equation}

Since $a$, $b$ and $c$ are functions of $\omega$ (from \eqref{eq:defabc}), these relations imply that there is a discrete set of frequencies $\{\omega_n\}$ for which the boundary conditions are satisfied. In the expressions below the `+ solutions' correspond to the quasinormal eigenfrequencies whereas the `$-$ solutions' correspond to the bound state eigenfrequencies. Additionally each of the modes have two types of eigenfrequencies, corresponding to the two possible relations in \eqref{eq:anb} and \eqref{eq:cacbn}. We denote these types by `right' and `left' frequencies, respectively, following the AdS/CFT-inspired terminology \cite{Birmingham:2001hc}, and below we follow the notation of \cite{Chen:2009rf}.

The `right frequencies' $(\omega_{\pm})_n^{(R)}$ are given by
\begin{equation}
(\omega_{\pm})_n^{(R)} = \frac{\nu^2+3}{d^2 \delta^2 - 3(\nu^2-1)} \left\{ -d\delta \left( \frac{4kd}{\nu^2+3} + i \left( n + \frac{1}{2} \right) \right) \pm i (e - i \, \text{sgn}(k) f)  \right\} \, ,
\label{eq:rightfrequency}
\end{equation}
where
\begin{gather}
d = \frac{1}{r_+ - r_-} \, , \qquad
\delta = 2\nu(r_+ + r_-) - 2 \sqrt{(\nu^2+3) r_+ r_-} \, , \label{eq:defddelta} \displaybreak[0]  \\
e = \sqrt{\frac{\sqrt{E^2+F^2}+E}{2}} \, , \qquad
f 	= \sqrt{\frac{\sqrt{E^2+F^2}-E}{2}} \, , \displaybreak[0] \\
E = \frac{1}{4} \left( 1 + \frac{4m^2}{\nu^2+3} \right) d^2 \delta^2 - 3(\nu^2-1) \left[ \frac{1}{4} \left( 1 + \frac{4m^2}{\nu^2+3} \right) + \left( \frac{4kd}{\nu^2+3} \right)^2 - \left( n + \frac{1}{2} \right)^2 \right] \, , \displaybreak[0] \\
F = - 3(\nu^2-1) \left( n + \frac{1}{2} \right) \frac{8kd}{\nu^2+3} \, .
\end{gather}
The `left frequencies' $(\omega_{\pm})_n^{(L)}$ are given by
\begin{equation}
(\omega_{\pm})_n^{(L)} = - i \left[ (2n+1) \nu \mp \sqrt{ 3(\nu^2-1) \left( n + \frac{1}{2} \right)^2 + \frac{\nu^2+3}{4} \left( 1 + \frac{4m^2}{\nu^2+3} \right)} \right] \, .
\label{eq:leftfrequency}
\end{equation}

Before discussing the stability of these field modes let us first make some important remarks. Since the `left eigenfrequencies' \eqref{eq:leftfrequency} have no real part, $\varpi$ in \eqref{eq:behaviouratinfinity} is real and hence there are no `left quasinormal modes' (in the sense that they do not have the expected outgoing wavelike behaviour at infinity). So only the `$-$ solution' is relevant for the `left modes'. Moreover, note that although there is a region of the parameter space for which the imaginary part of the `right quasinormal frequencies' \eqref{eq:rightfrequency} is positive it is easy to check that in this case either the mode is not outgoing at infinity or it decreases exponentially at infinity and hence is not a quasinormal mode. Otherwise, both the quasinormal and bound state frequencies have negative imaginary part and therefore these modes are classically stable.

It should also be noted that the bound state modes presented here are called quasinormal modes in some of the literature \cite{Oh:2008tc,Chen:2009rf,Chen:2009hg}. This is due to the adoption of different boundary conditions at infinity, motivated by AdS/CFT purposes \cite{Konoplya:2011qq,Berti:2009kk}. In fact, in the BTZ limit $\nu \to 1$ the bound state frequencies reduce to the quasinormal frequencies of the BTZ black hole in a rotating frame \cite{Cardoso:2001hn,Birmingham:2001hc}. This is expected since the the BTZ black hole quasinormal modes must vanish at infinity.

We conclude that the spacelike stretched black hole is classically stable to massive scalar field perturbations. In particular, there are no superradiant instabilities, even though superradiant modes can exist in this spacetime, as we described in section \ref{sec:superradiance}. In fact, it is straightforward to show using \eqref{eq:rightfrequency}, \eqref{eq:leftfrequency} and \eqref{eq:omegaH} that the bound state eigenfrequencies do not satisfy the superradiance condition $0 < \omega_{\text{R}} < k \Omega_{\mathcal{H}}$. This is related to the fact that the effective potential $V_{\omega k}(r)$ does not have a potential well where these superradiant modes could be localised, as illustrated in the plots of Fig.~\ref{fig:potentialmass}. The absence of superradiant instabilities is the main conceptual difference in classical scalar field theory between the spacelike stretched black hole and the Kerr spacetime \cite{Cardoso:2004nk,Cardoso:2005vk,Dolan:2007mj,Pani:2012vp,Dolan:2012yt,Witek:2012tr}.

\begin{figure}[t!]
\begin{center}
\subfigure[\; Quasinormal frequencies.]{
\includegraphics[scale=0.41]{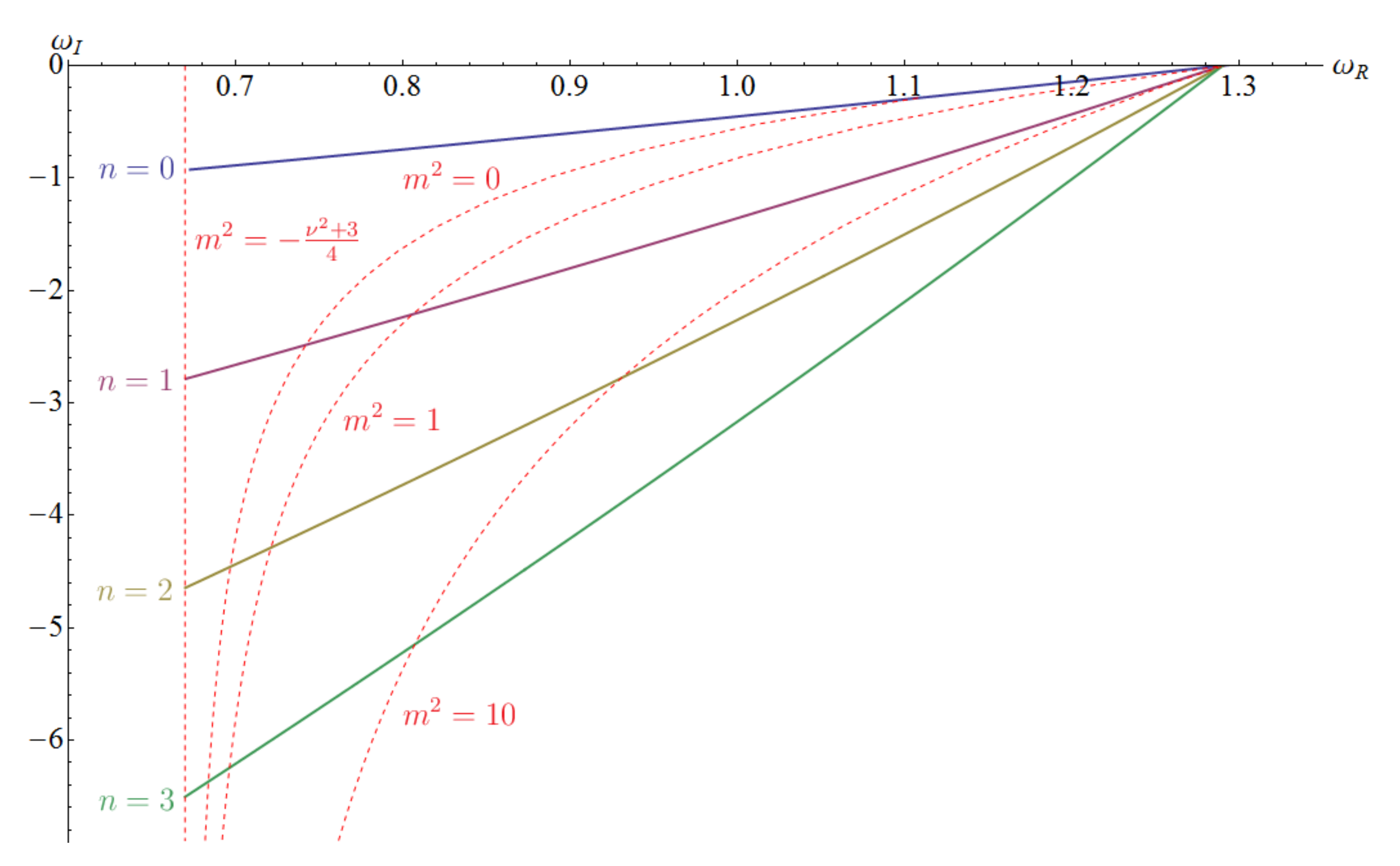}}
\subfigure[\; Bound state frequencies.]{
\includegraphics[scale=0.41]{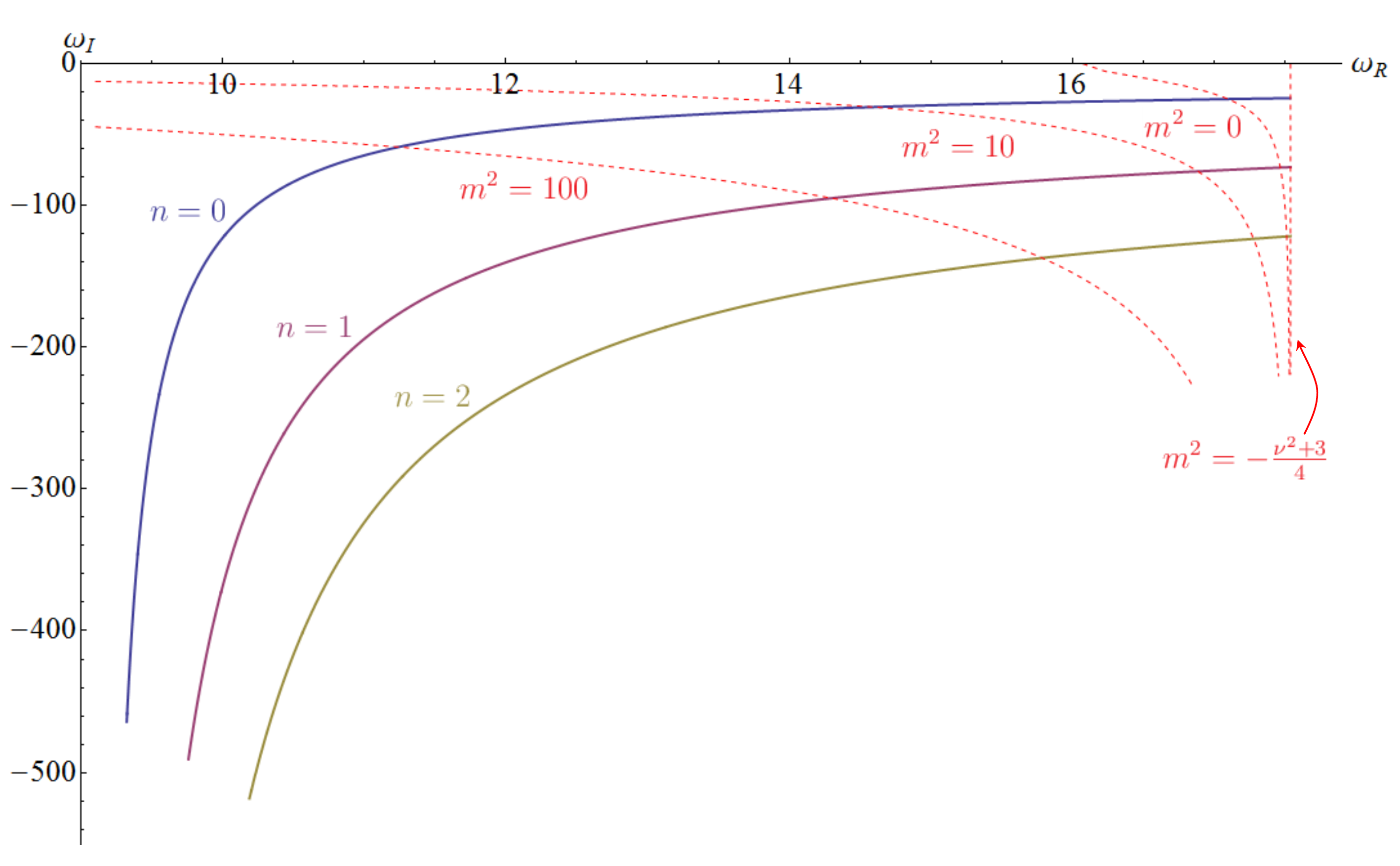}}
\caption{\label{fig:complexplane} Eigenfrequencies in the complex plane for a spacelike stretched black hole with $r_+ = 5$, $r_- = 0.5$ and $\nu=1.2$ and a scalar field with $k = -1$ and varying $m^2$. The different solid lines represent the eigenfrequencies for different overtone numbers $n$ and the dotted lines are lines of constant $m^2$. For a scalar field with a given $m^2$ there is a discrete set of complex eigenfrequencies for both the quasinormal and bound state modes at the intersection of the dotted and solid curves. If one lets $m^2$ run from its minimum possible value $-\frac{\nu^2+3}{4}$ one obtains these plots of the eigenfrequencies as functions of $m^2$. In the case of the quasinormal modes, for each $n$ there is a maximum value of $m^2$ after which there are no quasinormal modes of overtone number $n$. As for the bound state modes, they exist for any value of $m^2 \geq -\frac{\nu^2+3}{4}$.}
\end{center}
\end{figure}

In Fig.~\ref{fig:complexplane} we plot the quasinormal and bound state frequencies in the complex plane as functions of $m^2$. It is clear the discrete nature of the allowed frequencies and the fact that their imaginary part is always negative. For the quasinormal modes the real part of the frequency increases as we consider scalar fields of larger $m^2$, while the imaginary part decreases. The bound $m^2 \geq -\frac{\nu^2+3}{4}$ is a consequence of the constraint on the effective potential at infinity, as discussed in section \ref{sec:superradiance}. For each overtone number $n$ there is a maximum value of $m^2$ beyond which the corresponding quasinormal mode ceases to exist. This behaviour is similar to that of a massive scalar field in Schwarzschild and Kerr spacetimes \cite{Simone:1991wn}, as expected. Finally, the real and imaginary parts of the bound state modes frequencies are generally larger in absolute value than those of the quasinormal modes, but no growing modes are present.

%% CASE WITH MIRROR

\subsection{Bound state modes with a mirror outside the horizon}
\label{sec:mirror}

As we saw above, massive scalar fields propagating in spacelike stretched black holes do not give rise to classical instabilities. In particular, there are no superradiant bound state modes as the effective potential never develops a potential well. We now investigate whether these properties persist when a mirror-like boundary is introduced outside the event horizon. One reason to consider this situation is that a `mirror wall' in the effective potential might give rise to a superradiant `black hole bomb' instability \cite{Press:1972zz}, as is shown to happen for a massless scalar field in Kerr \cite{Cardoso:2004nk}, even though there are no superradiant instabilities when no mirror is present. Another reason is that if one wishes to quantise the scalar field, the existence of a speed-of-light surface implies that there is no well-defined Hartle-Hawking-like vacuum \cite{Ottewill:2000qh,Duffy:2005mz}. One way to solve this problem is precisely to add a mirror between the horizon and the speed-of-light surface.

\begin{figure}[t!]
\begin{center}
\subfigure[\; Real part of $\omega_{\mathcal{M}}$.]{
\includegraphics[scale=0.34]{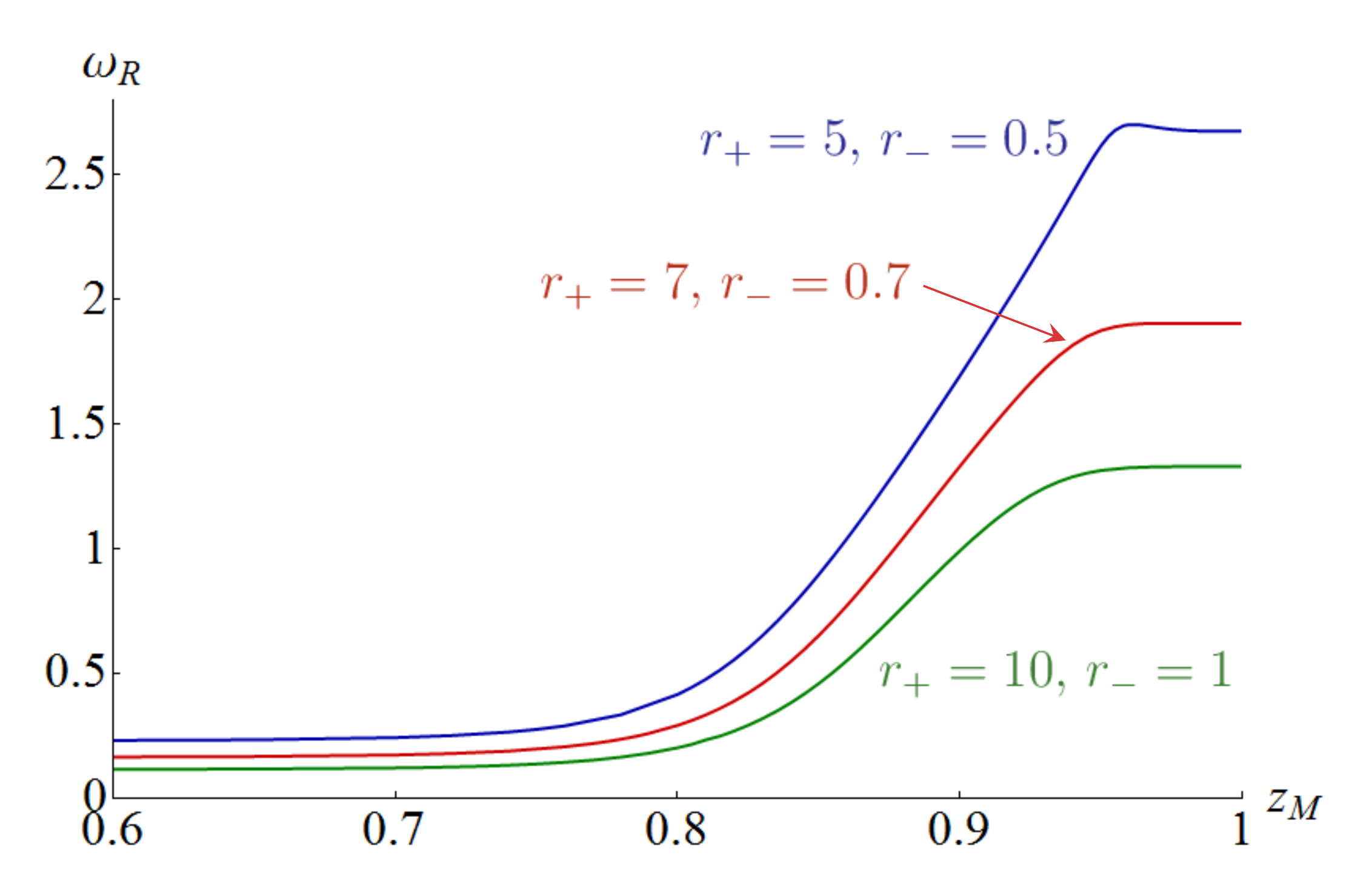}}
\subfigure[\; Imaginary part of $\omega_{\mathcal{M}}$.]{
\includegraphics[scale=0.34]{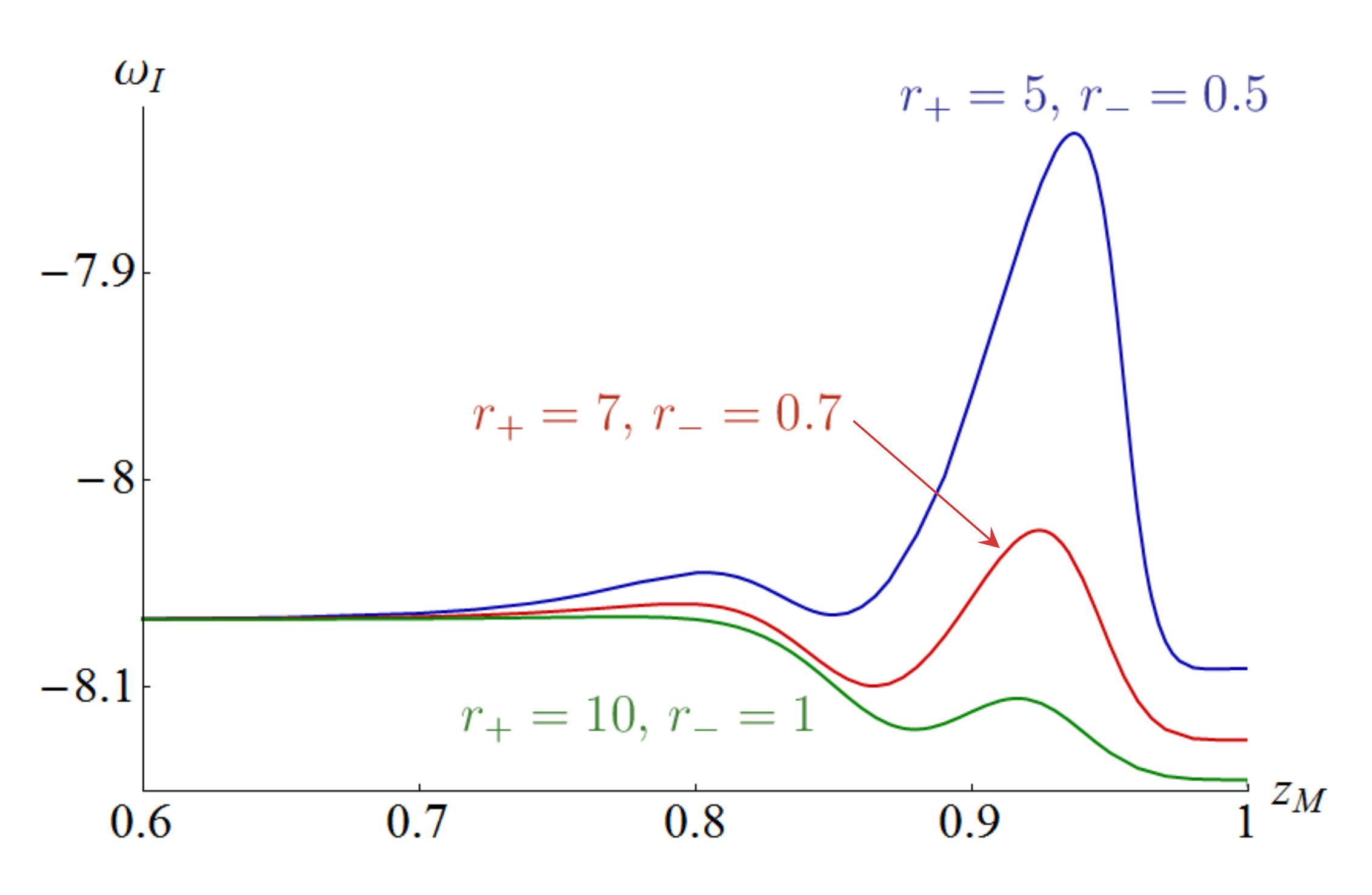}}
\caption{\label{fig:mirrorsize} `Right eigenfrequencies' as functions of the mirror's location for selected values of $r_+$ and $r_-$, with fixed $\nu=1.2$, $n=0$, $k=-1$ and $m=0$. The real part of the frequency is such that $\omega_{\text{R}} \to \text{Re} \big[(\omega_-)_0^{(R)} \big]$ in \eqref{eq:rightfrequency} as $z_{\mathcal{M}} \to 1$ and $\omega_{\text{R}} \to k \Omega_{\mathcal{H}}$ as $z_{\mathcal{M}} \to 0$. $k \Omega_{\mathcal{H}}$ equals 0.2307, 0.1648 and 0.1154 in the cases $r_+ = 5$, $r_+=7$ and $r_+=10$, respectively. The imaginary part of the frequency is such that $\omega_{\text{I}} \to \text{Im} \big[(\omega_-)_0^{(R)} \big]$ in \eqref{eq:rightfrequency} as $z_{\mathcal{M}} \to 1$.}
\end{center}
\end{figure}

Suppose then that a mirror-like boundary $\mathcal{M}$ is introduced at the radius $r=r_{\mathcal{M}}$, such that $r_+ < r_{\mathcal{M}} < \infty$. We impose Dirichlet boundary conditions at the mirror,
\begin{equation}
\Phi(t,z_{\mathcal{M}},\theta) = A_{\omega k} \, e^{-i\omega t + i k \theta} z^{\alpha}_{\mathcal{M}} (1-z_{\mathcal{M}})^{\beta} F(a,b,c;z_{\mathcal{M}}) = 0 \, ,
\end{equation}
where $z_{\mathcal{M}} = (r_{\mathcal{M}} - r_+)/(r_{\mathcal{M}} - r_-)$, according to \eqref{eq:definitionz}. The bound state modes now have eigenfrequencies $\omega_{\mathcal{M}}$ determined by:
\begin{equation}
z^{\alpha(\omega_{\mathcal{M}})}_{\mathcal{M}} (1-z_{\mathcal{M}})^{\beta(\omega_{\mathcal{M}})} F(a(\omega_{\mathcal{M}}),b(\omega_{\mathcal{M}}),c(\omega_{\mathcal{M}});z_{\mathcal{M}}) = 0 \, .
\end{equation}
Unfortunately, this equation cannot be analytically solved for $\omega_{\mathcal{M}}$, so we find these eigenfrequencies numerically, by truncating the hypergeometric series to the desired accuracy and using Mathematica's root finding algorithm. A check on the numerics is done by considering the limit $z_{\mathcal{M}} \to 1$ ($r_{\mathcal{M}} \to + \infty$), in which $\omega_{\mathcal{M}}$ approaches the previously derived bound state frequency $\omega_-$ without the mirror. Then we can use continuity to obtain the eigenfrequencies for any value of $z_{\mathcal{M}} \in (0,1)$.

Even though we do not have an explicit expression for the frequencies it is possible to obtain a useful piece of information by using the following heuristic argument. On the one hand one can only expect superradiant instabilities if the frequencies of the bound state modes are such that $\omega_R \lesssim \Omega_{\mathcal{H}}$ or, in other words, if their wavelengths are $\lambda \gtrsim \Omega_{\mathcal{H}}^{-1}$. On the other hand a mirror at $r = r_{\mathcal{M}}$ can only `see' these modes if $r_{\mathcal{M}} \gtrsim \lambda \gtrsim \Omega_{\mathcal{H}}^{-1}$. Therefore, superradiant instabilities, if they exist, can only occur if the mirror is placed beyond a critical radius which depends on the parameters of the spacetime. If one is not able to find any instabilities beyond this critical radius then one can assert with confidence that there are no superradiant instabilities wherever the mirror-like boundary is placed.

\begin{figure}[t!]
\begin{center}
\subfigure[\; Real part of $\omega_{\mathcal{M}}$.]{
\includegraphics[scale=0.34]{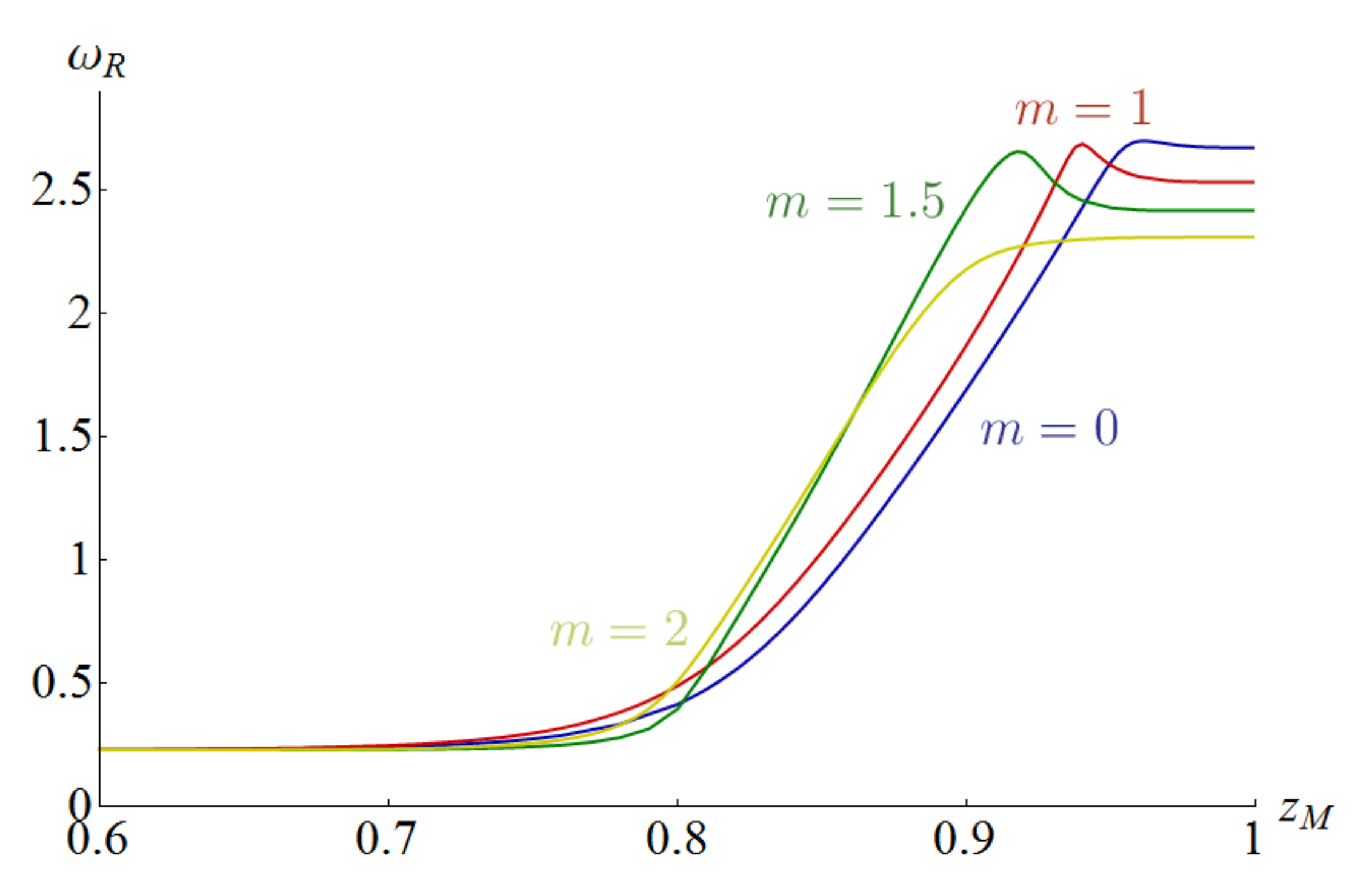}} \\
\subfigure[\; Imaginary part of $\omega_{\mathcal{M}}$.]{
\includegraphics[scale=0.34]{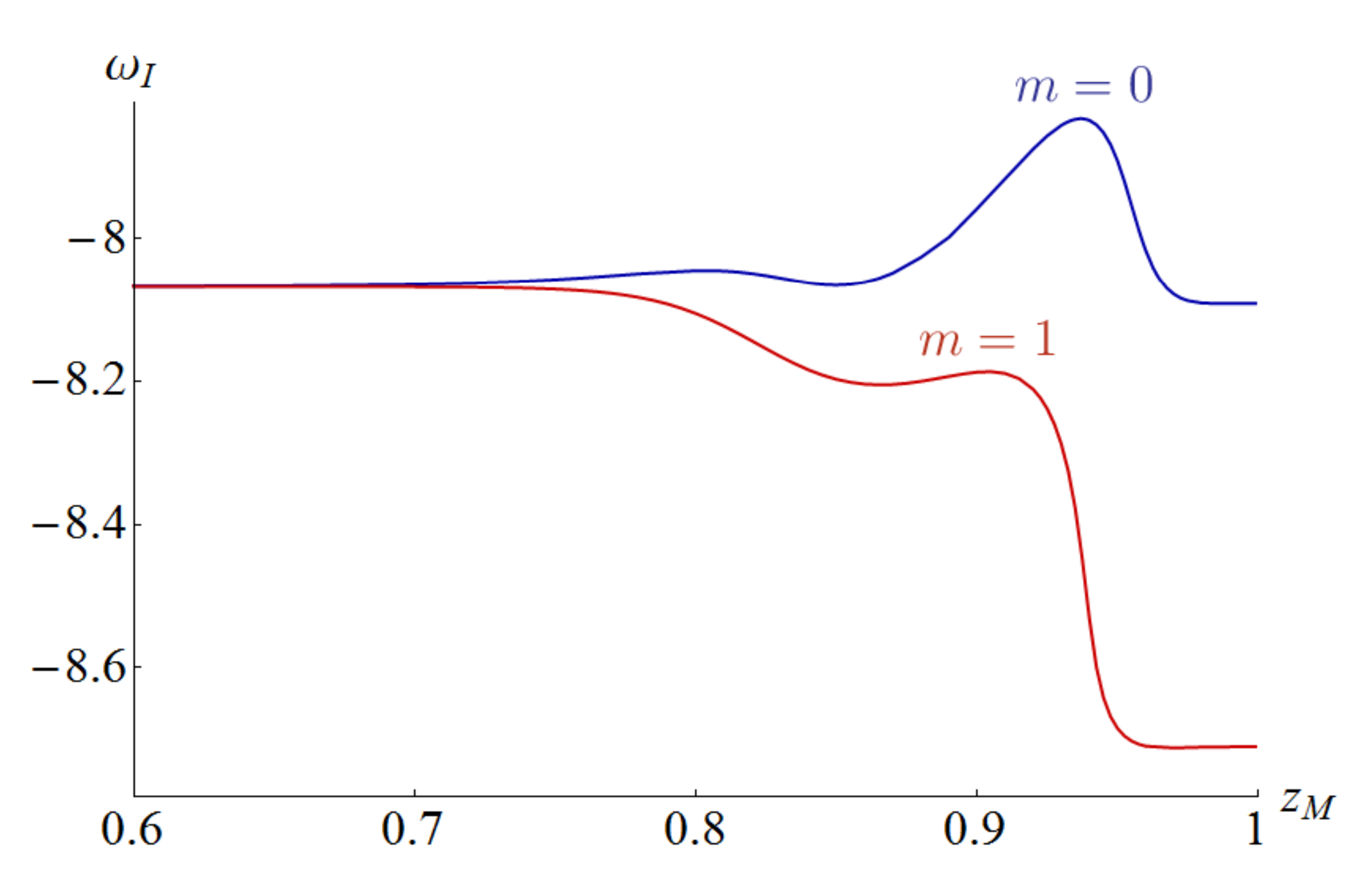}}
\subfigure[\; Imaginary part of $\omega_{\mathcal{M}}$.]{
\includegraphics[scale=0.34]{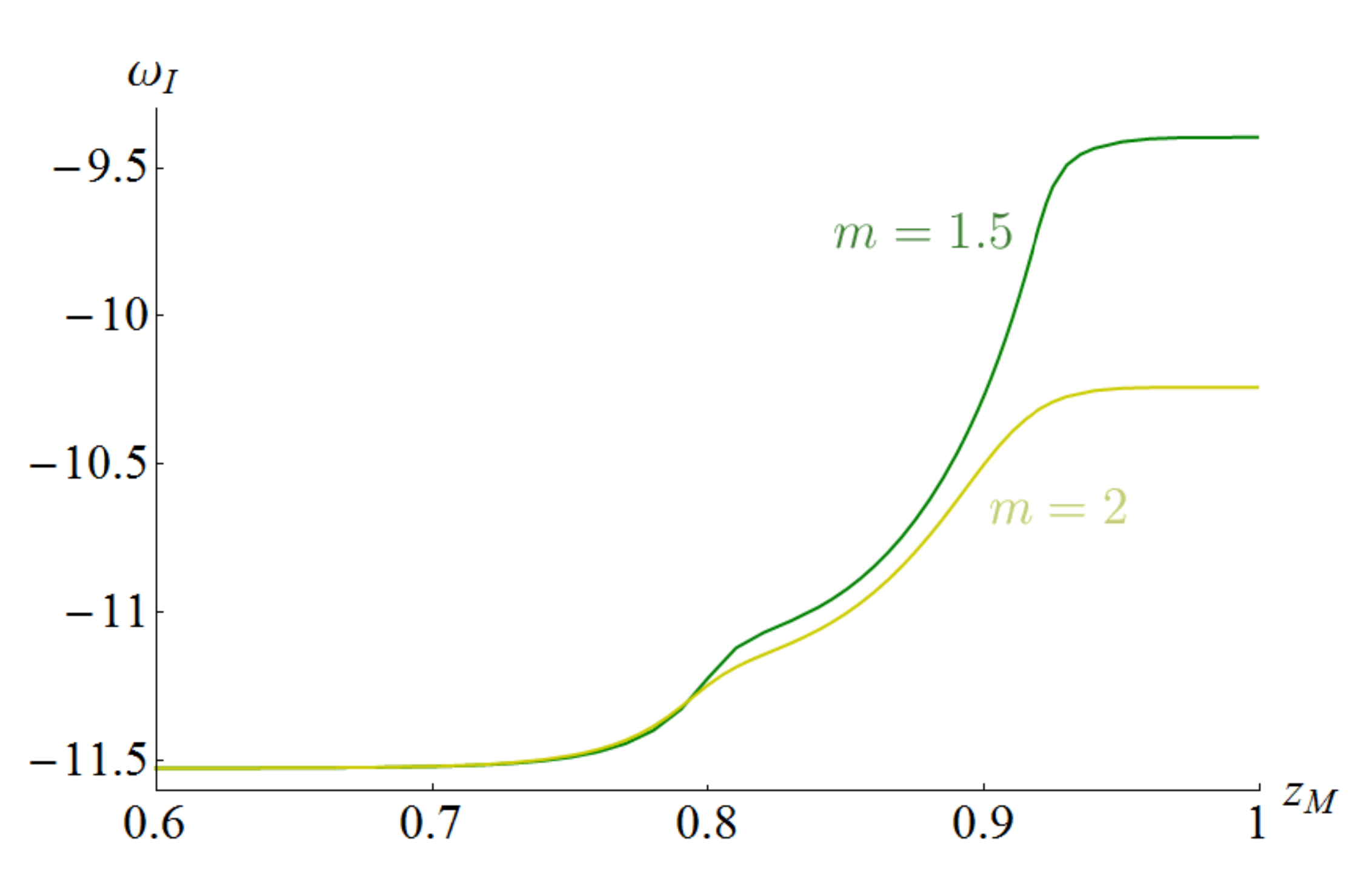}}
\caption{\label{fig:mirrormass} `Right eigenfrequencies' as functions of the mirror's location for selected values of $m$, with fixed $r_+ = 5$, $r_- = 0.5$, $\nu=1.2$, $n=0$ and $k = -1$. The real part of the frequency is such that $\omega_{\text{R}} \to \text{Re} \big[(\omega_-)_0^{(R)} \big]$ in \eqref{eq:rightfrequency} as $z_{\mathcal{M}} \to 1$ and $\omega_{\text{R}} \to k \Omega_{\mathcal{H}} = 0.2307$ as $z_{\mathcal{M}} \to 0$. The imaginary part of the frequency is such that $\omega_{\text{I}} \to \text{Im} \big[(\omega_-)_0^{(R)} \big]$ in \eqref{eq:rightfrequency} as $z_{\mathcal{M}} \to 1$.}
\end{center}
\end{figure}
\begin{figure}[t!]
\begin{center}
\subfigure[\; Real part of $\omega_{\mathcal{M}}$.]{
\includegraphics[scale=0.34]{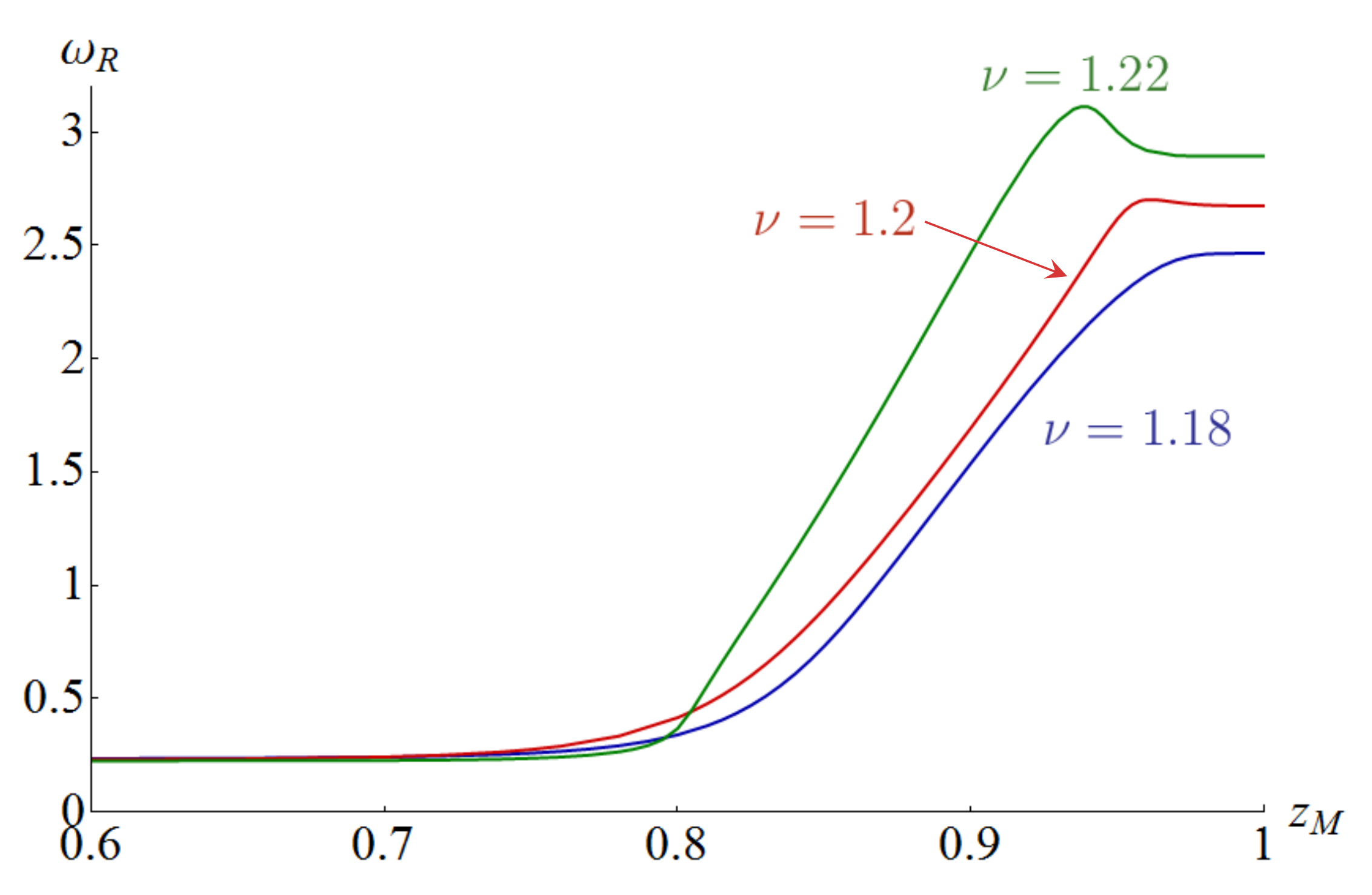}} \\
\subfigure[\; Imaginary part of $\omega_{\mathcal{M}}$.]{
\includegraphics[scale=0.34]{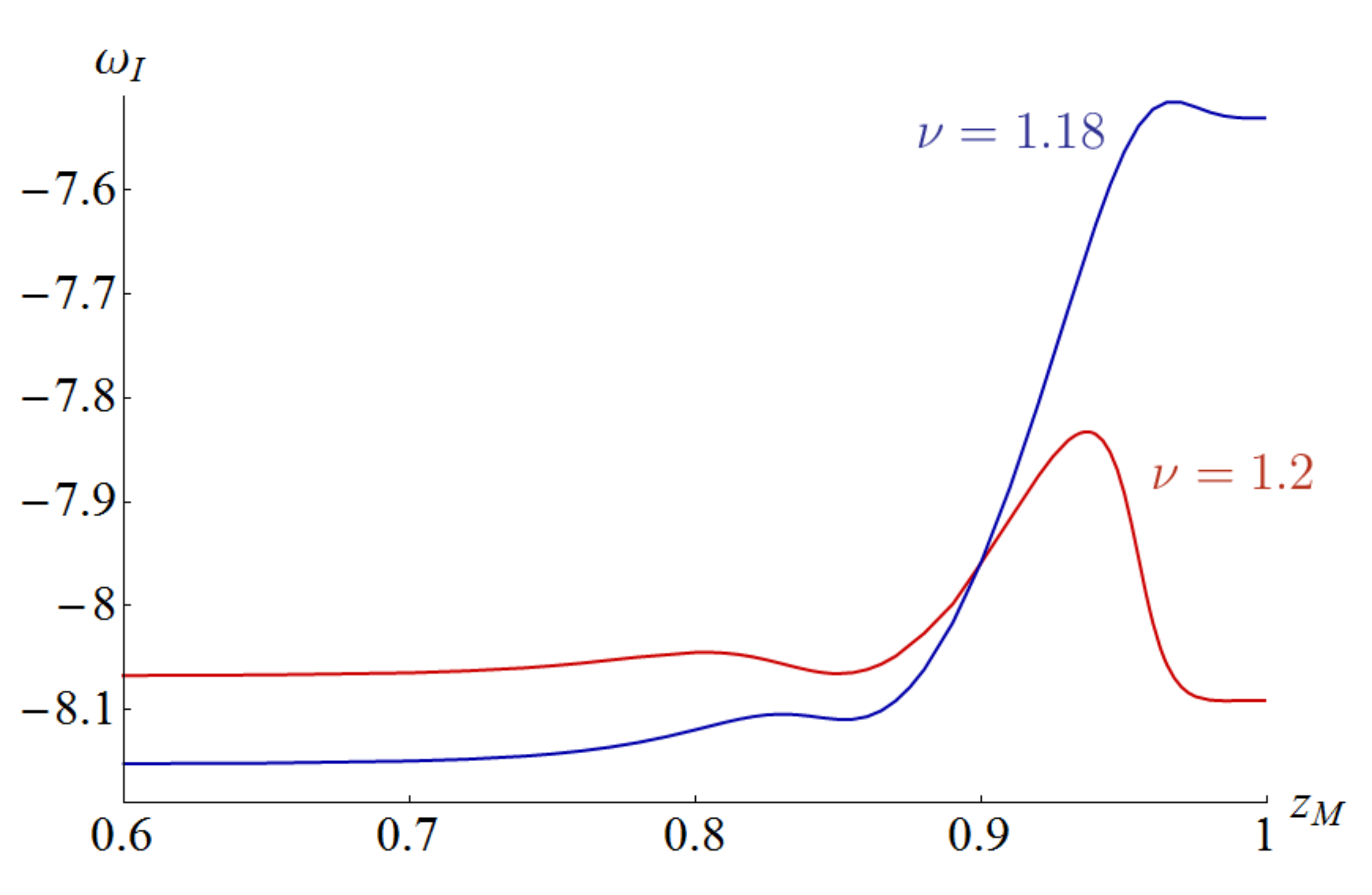}}
\subfigure[\; Imaginary part of $\omega_{\mathcal{M}}$.]{
\includegraphics[scale=0.34]{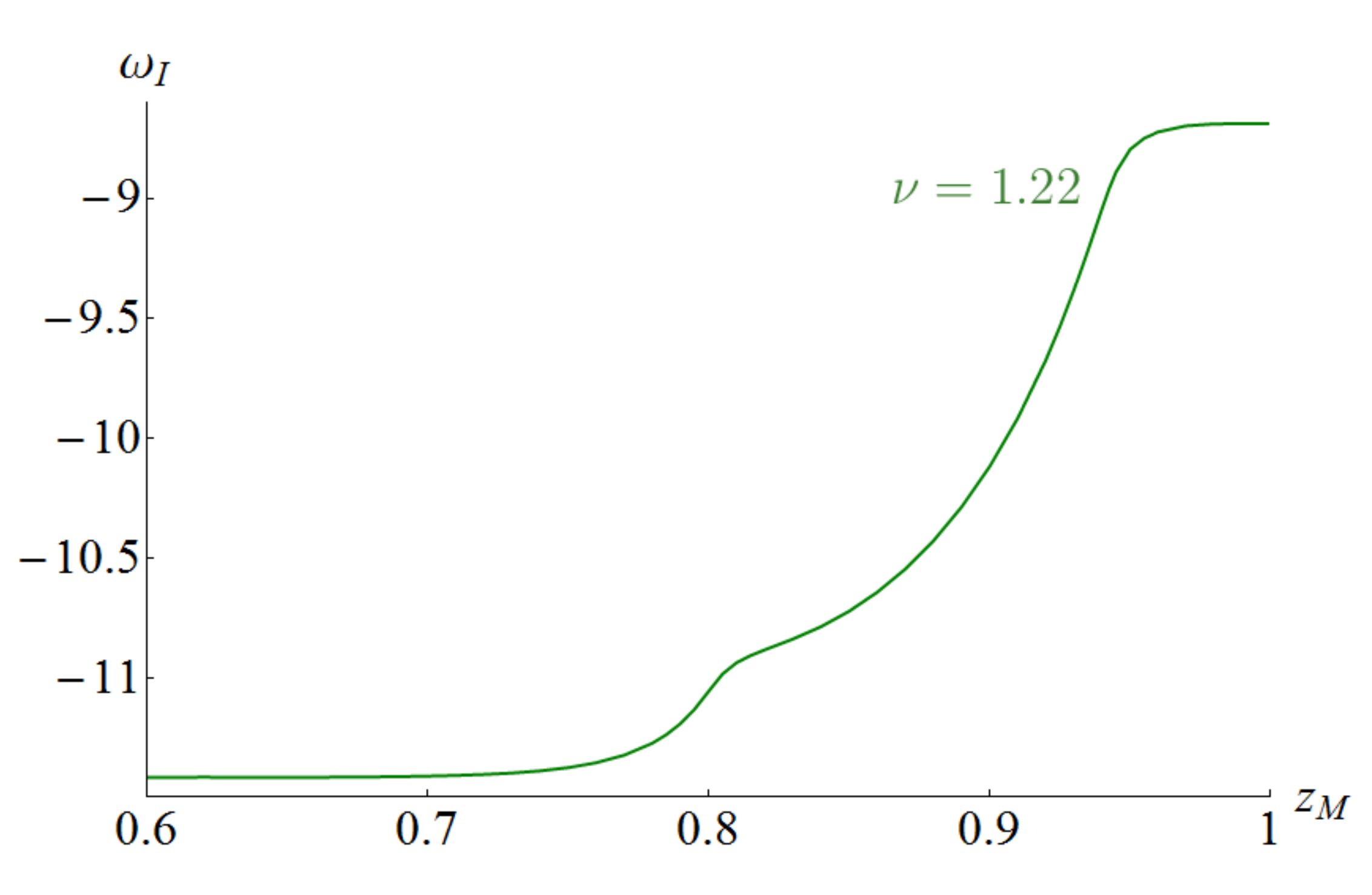}}
\caption{\label{fig:mirrornu} `Right eigenfrequencies' as functions of the mirror's location for selected values of $\nu$, with fixed $r_+ = 5$, $r_- = 0.5$, $n=0$, $k = -1$ and $m=0$. The real part of the frequency is such that $\omega_{\text{R}} \to \text{Re} \big[(\omega_-)_0^{(R)} \big]$ in \eqref{eq:rightfrequency} as $z_{\mathcal{M}} \to 1$ and $\omega_{\text{R}} \to k \Omega_{\mathcal{H}}$ as $z_{\mathcal{M}} \to 0$. $k \Omega_{\mathcal{H}}$ equals 0.2357, 0.2307 and 0.2260 in the cases $\nu=1.18$, $\nu=1.2$ and $\nu=1.22$, respectively. The imaginary part of the frequency is such that $\omega_{\text{I}} \to \text{Im} \big[(\omega_-)_0^{(R)} \big]$ in \eqref{eq:rightfrequency} as $z_{\mathcal{M}} \to 1$.}
\end{center}
\end{figure}
\begin{figure}[t!]
\begin{center}
\subfigure[\; Real part of $\omega_{\mathcal{M}}$.]{
\includegraphics[scale=0.34]{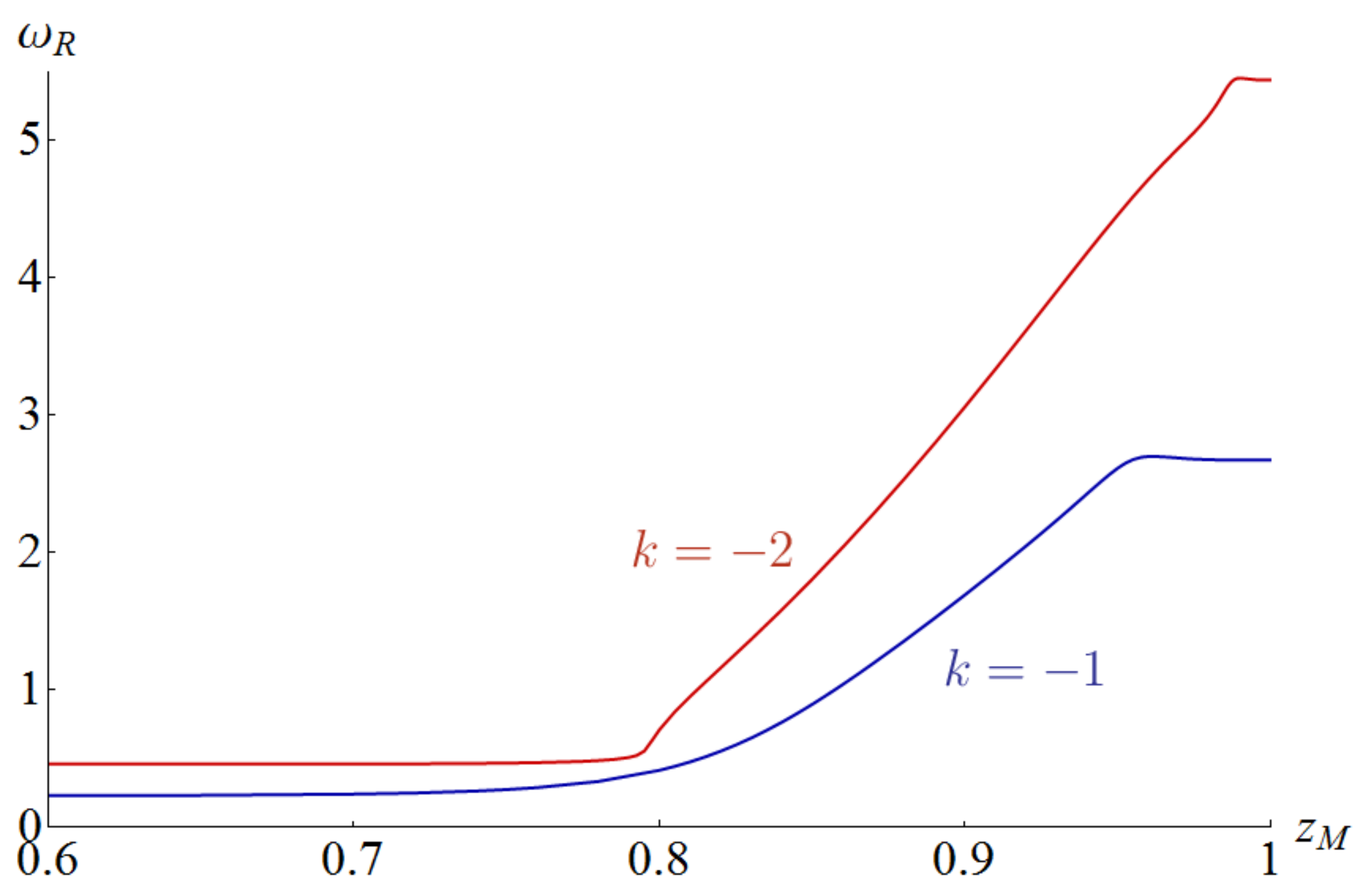}}
\subfigure[\; Imaginary part of $\omega_{\mathcal{M}}$.]{
\includegraphics[scale=0.34]{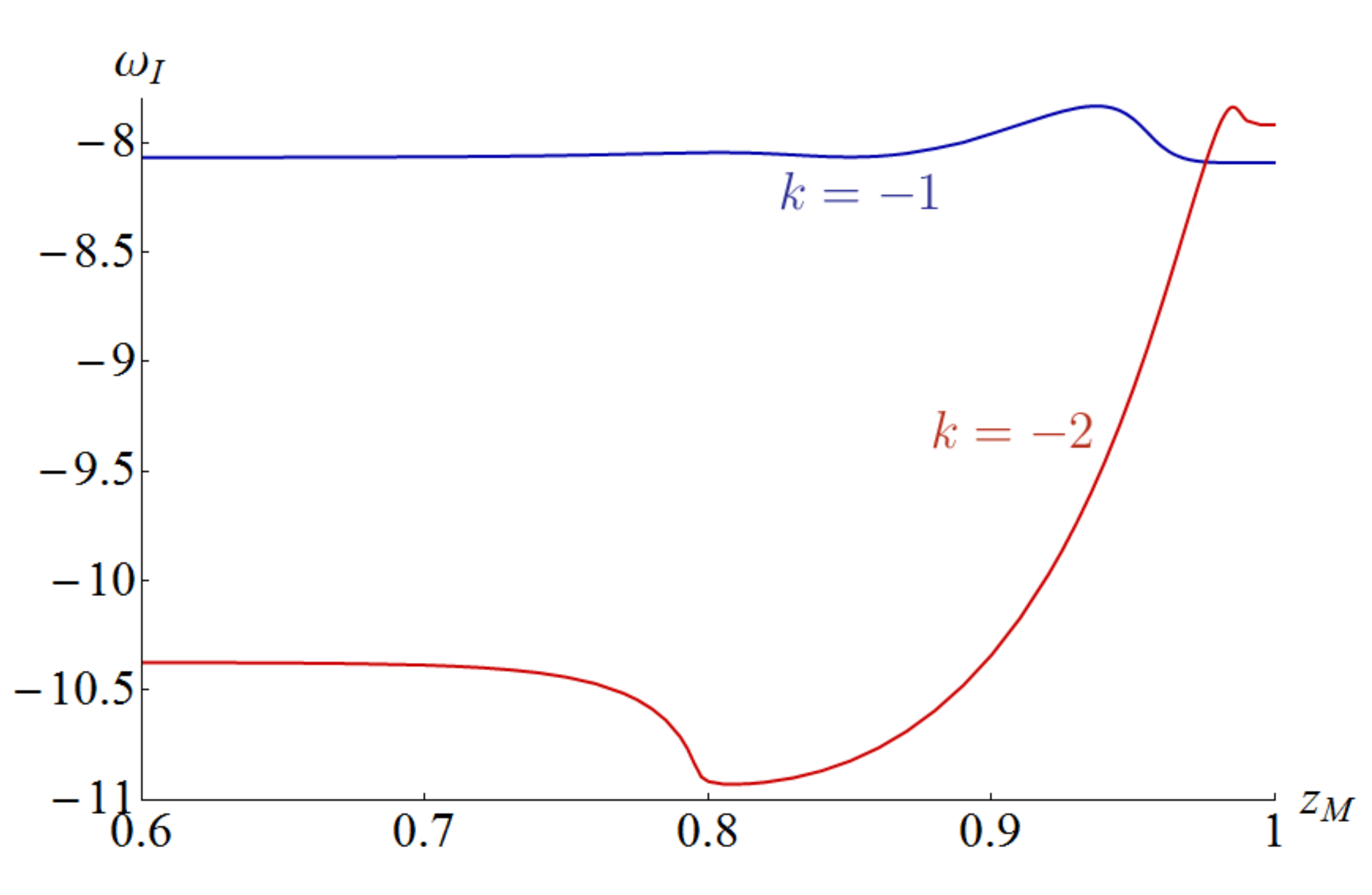}}
\caption{\label{fig:mirrork} `Right eigenfrequencies' as functions of the mirror's location for selected values of $k$, with fixed $r_+ = 5$, $r_- = 0.5$, $\nu=1.2$, $n=0$ and $m=0$. The real part of the frequency is such that $\omega_{\text{R}} \to \text{Re} \big[(\omega_-)_0^{(R)} \big]$ in \eqref{eq:rightfrequency} as $z_{\mathcal{M}} \to 1$ and $\omega_{\text{R}} \to k \Omega_{\mathcal{H}}$ as $z_{\mathcal{M}} \to 0$. $k \Omega_{\mathcal{H}}$ equals 0.2307 and 0.4615 in the cases $k=-1$ and $k=-2$, respectively. The imaginary part of the frequency is such that $\omega_{\text{I}} \to \text{Im} \big[(\omega_-)_0^{(R)} \big]$ in \eqref{eq:rightfrequency} as $z_{\mathcal{M}} \to 1$.}
\end{center}
\end{figure}
\begin{figure}[t!]
\begin{center}
\subfigure[\; Real part of $\omega_{\mathcal{M}}$.]{
\includegraphics[scale=0.34]{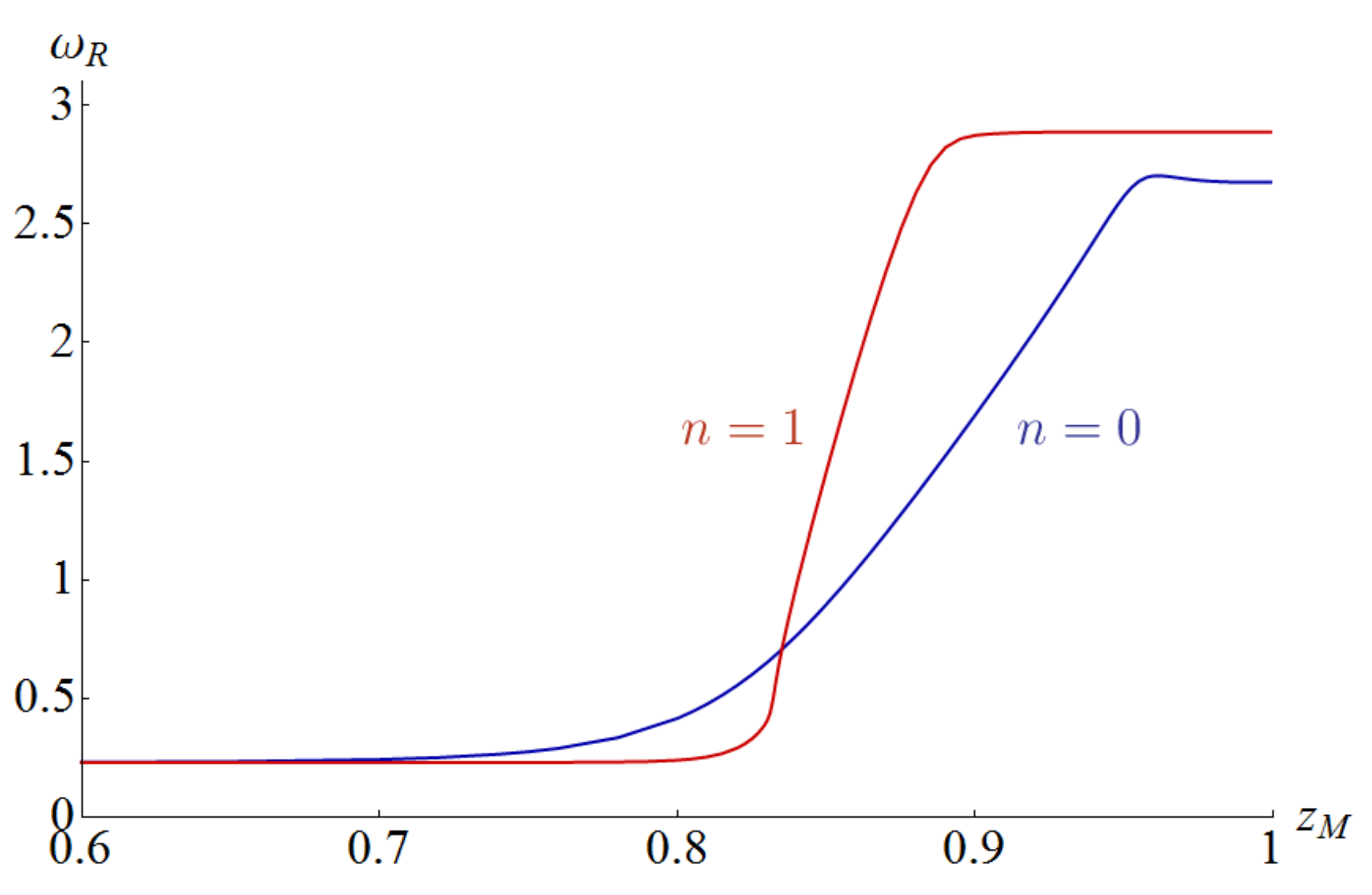}}
\subfigure[\; Imaginary part of $\omega_{\mathcal{M}}$.]{
\includegraphics[scale=0.34]{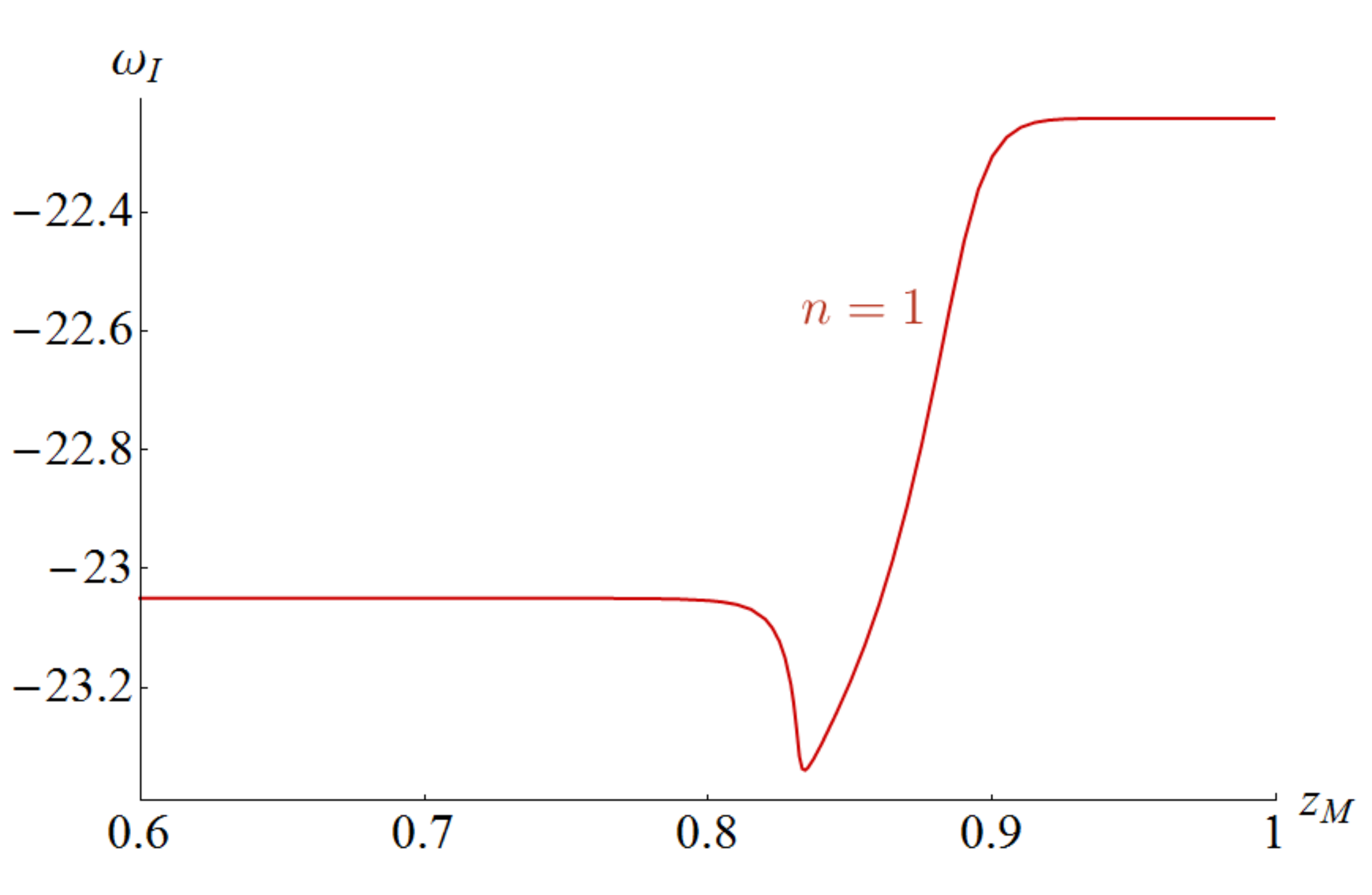}}
\caption{\label{fig:mirrorn} `Right eigenfrequencies' as functions of the mirror's location for selected values of $n$, with fixed $r_+ = 5$, $r_- = 0.5$, $\nu=1.2$, $k=-1$ and $m=0$. The real part of the frequency is such that $\omega_{\text{R}} \to \text{Re} \big[(\omega_-)_n^{(R)} \big]$ in \eqref{eq:rightfrequency} as $z_{\mathcal{M}} \to 1$ and $\omega_{\text{R}} \to k \Omega_{\mathcal{H}} = 0.2307$ as $z_{\mathcal{M}} \to 0$. The imaginary part of the frequency is such that $\omega_{\text{I}} \to \text{Im} \big[(\omega_-)_n^{(R)} \big]$ in \eqref{eq:rightfrequency} as $z_{\mathcal{M}} \to 1$.}
\end{center}
\end{figure}

In figures \ref{fig:mirrorsize}, \ref{fig:mirrormass}, \ref{fig:mirrornu}, \ref{fig:mirrork} and \ref{fig:mirrorn} we present the results for the real and imaginary parts of the `right' eigenfrequencies $\omega_{\mathcal{M}}$ as functions of the mirror's position for selected values of the parameters. Note that only negative values of $k$ are considered in these examples since, by \eqref{eq:rightfrequency}, $(\omega_-)_n^{(R)}(k) = [(\omega_-)_n^{(R)}(-k)]^*$ and thus it suffices to consider modes with $\omega_R \geq 0$.

First, note that the imaginary part of the eigenfrequencies is negative in all the presented cases and therefore no superradiant instabilities are present. This again contrasts with the Kerr spacetime surrounded by a mirror, where even a massless scalar field has superradiant instabilities \cite{Cardoso:2004nk}. 

In regard to the size of the black hole, from figure \ref{fig:mirrorsize} one observes that the real part of the frequency generally decreases as the horizon grows, while the imaginary part of the frequency increases in absolute value. The dependence on the scalar field mass as shown in figure \ref{fig:mirrormass} is more complicated but it is clear that the imaginary part of the frequency also increases in absolute value as the field mass increases. A similar conclusion can be drawn from figures \ref{fig:mirrornu}, \ref{fig:mirrork} and \ref{fig:mirrorn} concerning the warp factor $\nu$, the angular momentum number $k$ (in absolute value) and the overtone number $n$.

In order to understand these results it is useful to keep in mind the effective potential picture described in section \ref{sec:superradiance}. Note that in the current situation the frequencies take imaginary values and hence this picture is not entirely accurate. Recall that the `right' bound state  frequencies \eqref{eq:rightfrequency} have a real part that always exceeds $k \Omega_{\mathcal{H}}$ and therefore there are no superradiant bound state modes when the mirror is placed far from the event horizon. As we saw previously this can be explained by the fact that the effective potential does not develop a potential well near the horizon where the field mode could be trapped. However, as we move the mirror closer to the horizon there is the possibility that a potential well can be artificially created since the mirror works as an infinite potential wall. If we place the mirror close to the horizon, the real part of the frequency is approximately $k \Omega_{\mathcal{H}}$, due to the dragging of the inertial frames. In the general case in which the mirror is somewhere in between the horizon and infinity we expect the real part of the frequency to be greater than $k \Omega_{\mathcal{H}}$ but smaller than the asymptotic value, with possibly an increasing profile as the mirror is moved towards infinity. This expectation is in good agreement with the numerical results. We thus conclude that the real part of the `right frequency' does not satisfy the superradiant condition irrespective of the mirror's position in the exterior region. One may ask why the mirror does not create an artificial potential well. The well might have been expected to arise in cases where the effective potential has a local maximum near the horizon and the mirror is placed close to the horizon. We find however that when the mirror approaches the maximum of the effective potential from the right, the real part of the frequency does not decrease quickly enough to create superradiant bound state modes. When the mirror is moved even closer to the horizon, the real part of the frequency has no other choice but to approach $k \Omega_{\mathcal{H}}$.

The dependence of the imaginary part of the frequency (and consequently on the decay rate) on the several parameters of the system can be interpreted in the same way. If one increases the absolute values of $m^2$, $\nu$, $k$ and $n$, the effective potential is changed in such a way that the local maximum tends to disappear (as in Fig.~\ref{fig:potentialmass}) and, as a result, the field is more stable. On the other hand the effective potential itself depends on the frequency of the field and in this case the previous behaviour roughly occurs if we decrease the real part of the frequency. 

\begin{figure}[t!]
\begin{center}
\subfigure[\; Real part of $\omega_{\mathcal{M}}$.]{
\includegraphics[scale=0.34]{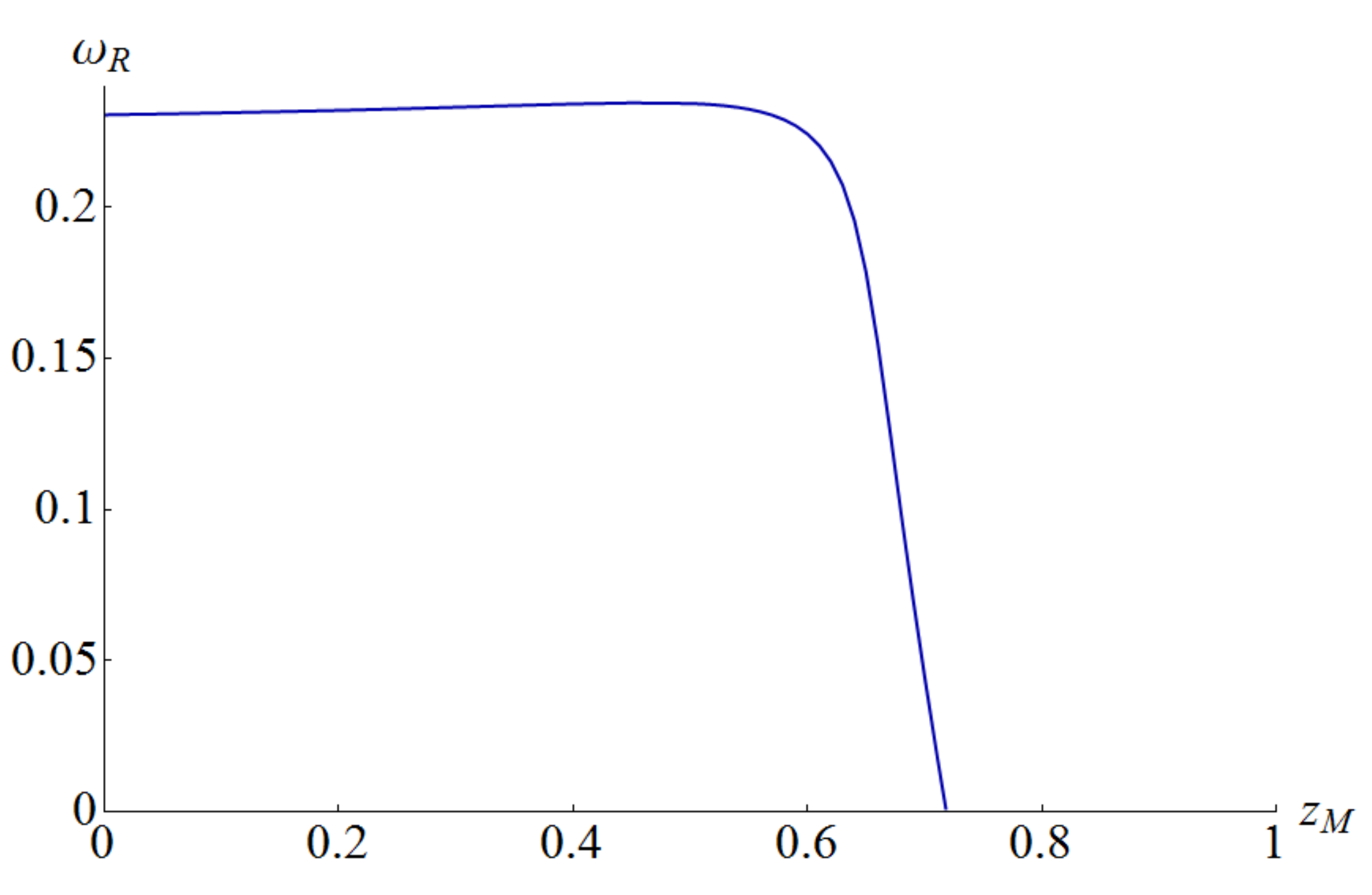}}
\subfigure[\; Imaginary part of $\omega_{\mathcal{M}}$.]{
\includegraphics[scale=0.34]{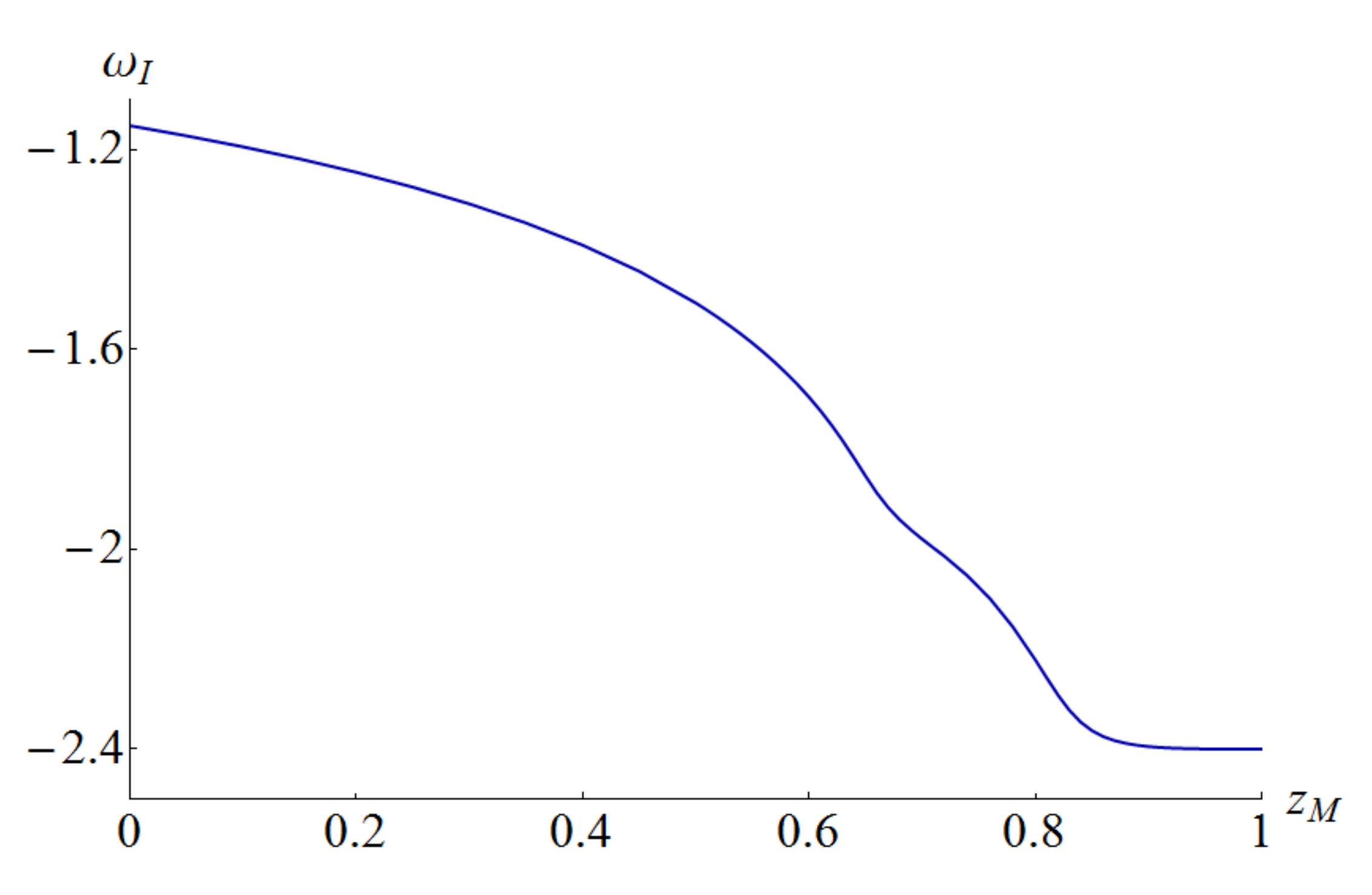}}
\caption{\label{fig:mirrorleft} `Left eigenfrequencies' as functions of the mirror's location for $r_+ = 5$, $r_- = 0.5$, $\nu=1.2$, $n=0$, $k=-1$ and $m=0$. The real part of the frequency is such that $\omega_{\text{R}} \to \text{Re} \big[(\omega_-)_0^{(L)} \big] = 0$ as $z_{\mathcal{M}} \to 1$ and $\omega_{\text{R}} \to k \Omega_{\mathcal{H}} = 0.2307$ as $z_{\mathcal{M}} \to 0$. The imaginary part of the frequency is such that $\omega_{\text{I}} \to \text{Im} \big[(\omega_-)_0^{(L)} \big]$ in \eqref{eq:rightfrequency} as $z_{\mathcal{M}} \to 1$.}
\end{center}
\end{figure}

The analysis for the `left eigenfrequencies' \eqref{eq:leftfrequency} has some similarities but also some significant differences. The dependence of the eigenfrequencies on the parameters of the system is largely identical. However, the dependence on the mirror's position is remarkably different. As we saw in section \ref{sec:qnm}, without the mirror the real part of the frequency is zero for the `left' bound state eigenfrequencies and so there is no superradiance. As seen in Fig.~\ref{fig:mirrorleft}, when the mirror is brought in from infinity this situation persists until a critical radius (call it $r_1$) beyond which the real part of the frequency sharply increases up to a value which is slightly greater than $k \Omega_{\mathcal{H}}$ (denote by $r_2$ the radius at which $\omega_{\text{R}} = k \Omega_{\mathcal{H}}$, such that $r_+ < r_2 < r_1$). When the mirror is placed at $r_{\mathcal{M}} \in (r_2,r_1)$ the bound state mode is indeed superradiant but the imaginary part of the eigenfrequency is still negative. Again, this can be understood by analysing the effective potential, which we recall depends on the frequency. The somewhat narrow interval of the mirror's position in which the real part of the frequency satisfies the superradiant condition is already past the local maximum of the effective potential, when it exists. Therefore, no potential well is created and thus no instabilities are present.

%%%%%%%%%%%%%%%%%%%%%%%%%%%%%%%%%%%%%%%%%%%%%%%%%%%%%%%%%%%%%%%%%%

%%% CONCLUSIONS

\section{Conclusions}
\label{sec:conclusions}

We have investigated classical properties of a massive scalar field on the background of a warped AdS${}_3$ black hole. The first main result is that classical superradiance is present when physically motivated boundary conditions are imposed at infinity; the second main result is that the black hole is nevertheless classically stable against the scalar perturbations, both with and without a stationary mirror in the exterior region. Taken together, these results contain an element of surprise as one might have expected the superradiant modes to create instabilities as in the (3+1)-dimensional Kerr spacetime. We showed however that instabilities are not present because the effective potential never develops a potential well near the horizon where the superradiant modes could be trapped. This stability might be a general characteristic of (2+1)-dimensional spacetimes and it is a particularly interesting result as almost all of the research to date on the classical stability of black holes has addressed spacetimes in four or more dimensions. Compare for instance with the results from \cite{Cardoso:2005vk} in which it was shown that black branes of the type Kerr${}_d \times \mathbb{R}^p$ (where $p \in \mathbb{N}$ and Kerr${}_d$ is the Kerr black hole if $d=4$ or the Myers-Perry black hole \cite{Myers:1986un} if $d>4$) have superradiant instabilities if $d=4$ but not if $d > 4$.

Additionally, this work clarifies the role of boundary conditions in classical superradiance. The `in' and `up' modes described in section \ref{sec:superradiance} can be superradiant, whatever the choice of positive frequency, similarly to what happens in the Kerr spacetime. This differs from the situation in the BTZ and Kerr-AdS spacetimes, where superradiance is not present if we choose reflective boundary conditions at infinity, which are motivated by their asymptotic structure. Also, we saw that superradiance can occur without a stationary limit surface, which would separate an ergoregion and the part of the exterior region connected to infinity. In our spacetime the ergoregion can be understood to extend to infinity since the Killing vector field $\partial_t$ is spacelike everywhere in the exterior region.

Given that the warped AdS${}_3$ black holes are classically stable to scalar field perturbations and that the existence of superradiance has been properly discussed, it is now possible to study quantum scalar field theory in these spacetimes. Some of the preliminary work has been done in this paper, namely we have constructed sets of basis field modes and introduced a mirror surrounding the black hole, which is the most straightforward way to avoid the speed-of-light surface, at which thermal states defined at the event horizon (such as the Hartle-Hawking vacuum) are divergent. We are encouraged by the technical simplicity provided by the lower dimensional setting. As an example if one wishes to find the renormalised expectation value of the stress-energy tensor in one of the thermal states of the scalar field, it should be possible to obtain the associated Green's function in terms of known functions, which may significantly reduce the required computational work.

Finally, it would also be interesting to extend this study to other kinds of perturbations, such as Proca and especially gravitational fields. The classical stability of BTZ black holes in TMG to gravitational perturbations has been established in \cite{Birmingham:2010gj}. A similar analysis for the warped AdS${}_3$ black hole is expected to be more involved, owing to the fewer symmetries of the spacetime, but the results of \cite{Anninos:2009zi} for the warped AdS${}_3$ vacuum solution suggests that this analysis may be doable. We hope to return to this point in the future.

%%%%%%%%%%%%%%%%%%%%%%%%%%%%%%%%%%%%%%%%%%%%%%%%%%%%%%%%%%%%%%%%%%

%%% ACKNOWLEDGMENTS

\begin{acknowledgements}
I would like to thank Jorma Louko and Vitor Cardoso for a critical reading of the manuscript and also Sam Dolan, Elizabeth Winstanley and Helvi Witek for helpful discussions and feedback. I acknowledge financial support from Funda\c{c}\~{a}o para a Ci\^{e}ncia e Tecnologia (FCT)-Portugal through Grant No. SFRH/BD/69178/2010.
\end{acknowledgements}

%%% BIBLIOGRAPHY

\end{document}